\documentclass[aps,prb,showpacs,showkeys,preprintnumbers,amsmath,amssymb,amsfonts,amsfontssuperscriptaddress,twocolumn]{revtex4}
\usepackage{graphicx}
\usepackage{epstopdf}
\usepackage{color}
\usepackage{accents}
\usepackage{bm}
\usepackage[colorlinks=true,linkcolor=blue,citecolor=red]{hyperref}%
\usepackage{url}
\usepackage{tikz-cd}
\usepackage{dsfont}
\usepackage{multirow}

\def\Pr{\mathcal{P}}

\def\A{\Gamma}

\def\L{\mathcal{L}}
\def\N{\mathcal{N}}
\def\P{\mathbb{P}}
\def\T{\mathcal{T}}

\def\Ti{\T^0}
\def\I{\mathbf{1}}

\def\x{\bm{x}}
\def\R{\mathbb{R}}

\def\koff{k_{\rm off}}
\def\kon{k_{\rm on}}
\def\pa{\partial\Omega}

\def\F{\mathcal{F}}

\def\Q{\mathcal{Q}}

\usepackage{epsfig}

\begin{document}

\title{Reversible Target-Binding Kinetics of Multiple Impatient Particles}

\author{Denis S. Grebenkov}
\affiliation{Laboratoire de Physique de la Mati\`ere Condens\'ee (UMR 7643), \\ CNRS -- Ecole Polytechnique, IP Paris, 91120 Palaiseau, France}
\email{denis.grebenkov@polytechnique.edu}

\author{Aanjaneya Kumar}
\affiliation{Department of Physics, Indian Institute of Science Education and Research, \\ Dr. Homi Bhabha Road, Pune 411008, India}
\email{kumar.aanjaneya@students.iiserpune.ac.in}

\date{\today}

\begin{abstract}
Certain biochemical reactions can only be triggered after binding of a
sufficient number of particles to a specific target region such as an
enzyme or a protein sensor.  We investigate the distribution of the
reaction time, i.e., the first instance when all independently
diffusing particles are bound to the target.  When each particle binds
irreversibly, this is equivalent to the first-passage time of the
slowest (last) particle.  In turn, reversible binding to the target
renders the problem much more challenging and drastically changes the
distribution of the reaction time.  We derive the exact solution of
this problem and investigate the short-time and long-time asymptotic
behaviors of the reaction time probability density.  We also analyze
how the mean reaction time depends on the unbinding rate and the
number of particles.  Our exact and asymptotic solutions are compared
to Monte Carlo simulations.
\end{abstract}



\maketitle

\section{Introduction}

Diffusion-controlled reactions play an important role in many chemical
and biological processes.  In a typical scenario, particles diffuse in
a confining domain towards a specific target region to react or to
trigger a biological event
\cite{Lauffenburger,Alberts,Redner,Schuss,Metzler,Oshanin,Grebenkov07,Benichou14,Holcman14}.
Various aspects of such diffusive search processes have been
investigated, with the particular emphasis on first-passage times
(FPTs) that characterize how fast a single particle finds a single
target.  The distribution of the first-passage time $\tau$ is usually
described by the survival probability, $S(t) = \P\{ \tau > t\}$, or,
equivalently, by the probability density function $H(t) = -dS(t)/dt$.
The distribution and, particularly, the mean value $\langle
\tau\rangle$ and the associate reaction rate, have been thoroughly
analyzed
\cite{Grigoriev02,Singer06a,Singer06b,Singer06c,Condamin07,Benichou08,Benichou10,Benichou10b,Pillay10,Cheviakov10,Grebenkov10,Cheviakov12,Caginalp12,Mattos12,Berezhkovsky12,Rupprecht15,Godec16,Godec16b,Grebenkov16,Marshall16,Grebenkov17b,Lanoiselee18,Grebenkov18a,Grebenkov18,Sposini19,Grebenkov19,Levernier19,Grebenkov19b,Hartich19,Hartich19b,Grebenkov20a,Grebenkov20b}.

As the diffusive search is typically long, many independent searchers
are generally involved to speed up this process.  In this setting, the
arrival of the {\it fastest} particle among $N$ particles can trigger
the reaction.  If $\tau_1,\ldots,\tau_N$ denote the FPTs of these
particles, the fastest first-passage time (fFPT) is $\Ti_{1,N} =
\min\{\tau_1,\ldots,\tau_N\}$.  As the particles search independently,
the distribution of the fFPT is simply
\begin{equation}  \label{eq:surv1}
\P\{\Ti_{1,N} > t \} = \P\{ \tau_1 > t\} \cdots \P\{ \tau_N > t\} = [S(t)]^N.  
\end{equation}
Similarly, the first-passage time $\Ti_{K,N}$ of the $K$-th fastest
particle to arrive onto the target is governed by the law
\begin{equation}
\P\{\Ti_{K,N} > t \} = \sum\limits_{j=0}^{K-1} {N \choose j} [S(t)]^{N-j} [1-S(t)]^{j} ,  
\end{equation}
where ${N \choose j}$ is the binomial coefficient.  The associated
probability density follows immediately:
\begin{eqnarray}  \nonumber
H_{K,N}^{0}(t) &=& - \frac{d\P\{\Ti_{K,N} > t \}}{dt} \\  \label{eq:HKN}
&=& K {N \choose K} [S(t)]^{N-K} [1-S(t)]^{K-1}  H(t).
\end{eqnarray}
While these expressions fully characterize the random variable
$\Ti_{K,N}$, finding the large-$N$ asymptotic behavior of its moments,
$\langle [\Ti_{K,N}]^m\rangle$, is a difficult problem.  More
generally, random variables $\Ti_{K,N}$ present an example of extreme
value statistics \cite{Majumdar20}.

This problem was first studied by Weiss {\it et al.} who showed by
analyzing the exact form of $S(t)$ for one-dimensional diffusion on an
interval that the mean $\langle \Ti_{K,N}\rangle$ decreases
logarithmically slowly with $N$: $\langle \Ti_{K,N}\rangle
\propto 1/\ln N$ as $N\to\infty$ \cite{Weiss83}.  They also briefly
considered higher-order moments and argued the universality of the
logarithmic decay for other diffusive processes.  This seminal work
was further extended by several authors
\cite{Basnayake18,Basnayake19,Lawley20a,Lawley20b,Reynaud15,Schuss19,Madrid20,Grebenkov20}.
For instance, Basnayake {\it et al.} as well as Lawley and Madrid gave
rigorous mathematical proofs for the asymptotic behavior of these
moments \cite{Basnayake18,Basnayake19,Lawley20a} (see also Appendix
\ref{sec:Tmean_irrev} for new results concerning the behavior of the
mean of the {\it slowest} FPT $\Ti_{N,N}$).  In addition, Lawley found
the parameters of the asymptotic Gumbel distribution of $\Ti_{1,N}$
for a large class of diffusion processes \cite{Lawley20b}.  Moreover,
this result was extended to the $K$-th fastest FPT $\Ti_{K,N}$ and the
asymptotic form of the joint distribution of $\{\Ti_{1,N},
\ldots, \Ti_{K,N}\}$ was derived.  The logarithmic scaling of the mean
fFPT was evoked to rationalize the redundancy in the number of
searchers in some biological systems, such as the large number of
sperm cells \cite{Reynaud15,Schuss19}.  We stress, however, that a
logarithmic speed up of the search process is too costly from a
practical point of view; for instance, a tenfold reduction of the mean
time would require more than twenty thousands of particles.  Note that
if the starting positions of $N$ particles are uniformly distributed
in the domain, the logarithmic decay is replaced by a much faster
power law decay: $\langle \Ti_{1,N}\rangle \propto 1/N$ for partially
reactive targets, and $\langle \Ti_{1,N}\rangle \propto 1/N^2$ for
perfectly reactive targets \cite{Grebenkov20}.  More generally, the
power law decay $1/N$ was shown to emerge as a transient regime of
moderately large $N$ if the target is small or if finding the target
requires escaping a potential well \cite{Madrid20}.  Moreover, as the
mean value is not always the relevant time scale of the process
\cite{Grebenkov17b,Grebenkov18,Reva21}, evolutionary optimization
of diffusive search does not necessarily aim at reducing the mean
value of the fFPT.

\begin{figure}
\begin{center}
\includegraphics[width=80mm]{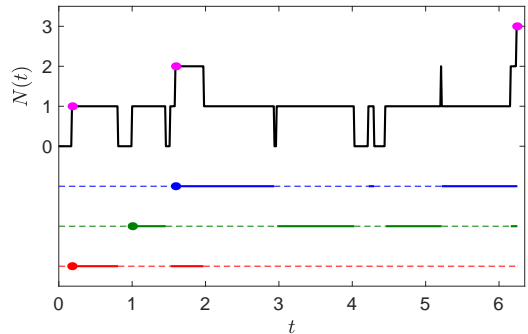} 
\end{center}
\caption{
Illustration of a simulated process $\N(t)$ that counts the number of
bound particles at time $t$, with $N = 3$.  Three filled circles
indicate the first-crossing times $\T_{K,N}$ for $K = 1,2,3$.  On the
bottom, there is a schematic presentation of the state of each of
three particles: free state (thin dashed line) vs bound state (thick
solid line).  Three filled circles indicate the first-passage times
$\Ti_{K,N}$ for the first, the second, and the third particles (in the
order of arrival).  While $\T_{1,N} = \Ti_{1,N}$, unbinding events
imply that $\T_{K,N} \geq \Ti_{K,N}$ for any $K > 1$.}
\label{fig:Nt}
\end{figure}

In the above discussion, each particle that arrived onto the target
was supposed to remain on it forever.  In particular, the number
$\N(t)$ of particles bound to the target at time $t$, is a
non-decreasing stochastic process that increases by $1$ at the arrival
of each new particle.  As a consequence, the $K$-th fastest FPT
$\Ti_{K,N}$ is equal to the first instance $\T_{K,N} = \inf\{ t > 0 ~:~
\N(t) = K\}$ when $K$ particles among $N$ are bound to the target.  In
many chemical and biological settings, however, binding to the target
is a {\it reversible} process, i.e., each particle remains on the
target for some waiting time, unbinds from it and resumes its bulk
diffusion.  The waiting time is usually considered to be an
independent random variable obeying an exponential law with the rate
$\koff$.  As each unbinding event diminishes $\N(t)$ by $1$, the
number of bound particles is no longer a non-decreasing process
(Fig. \ref{fig:Nt}).  Even though the dynamics of all particles is
Markovian (i.e., their positions and states at time $t$ fully
determine the probabilities of their positions and states in the
future), the number $\N(t)$ of bound particles is a non-Markovian
process.  The reversible binding does not affect the statistics of the
first instance $\T_{1,N}$ when one particle (the fastest one) is bound
to the target, i.e., Eq. (\ref{eq:surv1}) governs the probability law
for $\T_{1,N} = \Ti_{1,N}$.  In contrast, the first instance
$\T_{K,N}$ for $K$ particles to be bound to the target is no longer
equal to the $K$-th fastest FPT $\Ti_{K,N}$.  Indeed, before the
binding of the $K$-th particle, some of the previously bound particles
can unbind, and thus $\T_{K,N} \geq \Ti_{K,N}$ (the superscript $0$
highlights that the first-passage times $\Ti_{K,N}$ correspond to
irreversible binding with $\koff = 0$).  Even though the particles are
independent, random waiting times spent by these particles on the
target render the characterization of the reaction times $\T_{K,N}$
much more challenging than that of $\Ti_{K,N}$.  As reversible binding
allows for some particles to leave the target before the arrival of
the others, they were termed {\it impatient} \cite{Grebenkov17}.  In
Ref. \cite{Grebenkov17}, the problem of two impatient particles
diffusing on an interval was mapped onto intermittent diffusion on a
square.  Solving the latter problem, the mean $\langle
\T_{2,2}\rangle$ was obtained, and the effect of reversible binding
was analyzed.  Even this basic case with two particles required
sophisticated analysis.

Lawley and Madrid proposed a remarkable approximation to the general
problem \cite{Lawley19}.  Assuming that the first-binding time and the
rebinding time (i.e., the random time between unbinding of a particle
from the target and its next rebinding) can be approximated by an
exponential random variable with some rate $\nu$, the number $\N(t)$
can be modeled by a Markovian birth-death process $\bar{\N}(t)$
between $N+1$ states of $0,1,2,\ldots,N$ bound particles:
\begin{equation}
\begin{tikzcd}[every arrow/.append style={shift left}]
 0 \arrow{r}{N\nu} & 1 \arrow{l}{\koff}  \arrow{r}{(N-1)\nu} & 2 \arrow{l}{2\koff}  
\quad \cdots \quad N-1 \arrow{r}{\nu} & N \arrow{l}{N\koff} 
\end{tikzcd}
\end{equation}
(here and throughout the text, bar denotes the quantities
corresponding to the Lawley-Madrid approximation).  Introducing an
$(N+1)\times(N+1)$-dimensional matrix $W$ with zero elements except
for
\begin{equation*}
W_{i,i+1} = i \koff,  \quad W_{i+1,i} = (N+1-i) \nu \quad (i=1,2,\ldots,N),
\end{equation*}
and $W_{i,i}$ are chosen so that $W$ has zero column sums, the
distribution of the first-crossing time $\bar{\T}_{K,N} = \inf\{ t>0
~:~ \bar{\N}(t) = K\}$ can be written as \cite{Lawley19}
\begin{equation}  \label{eq:SKN_LMA}
\P\{ \bar{\T}_{K,N} > t\} = \sum\limits_{j=1}^K \bigl[\exp(W^{(K)} t)\bigr]_{j,1} \,,
\end{equation}
where $W^{(K)}$ is the $K\times K$ matrix obtained by retaining the
first $K$ columns and $K$ rows from $W$ and discarding everything
else.  Here the initial state was assumed to be $0$, i.e., no bound
particles.  In other words, the distribution is expressed in terms of
the matrix exponential of $W^{(K)}$.  The probability density of
$\bar{\T}_{K,N}$ is even simpler:
\begin{equation}  \label{eq:HKN_LMA}
\bar{H}_{K,N}(t) = \nu (N-K+1) \bigl[\exp(W^{(K)} t)\bigr]_{K,1} \,.
\end{equation}
Finally, the mean time is fully explicit:
\begin{equation}  \label{eq:LMA_mean}
\langle \bar{\T}_{K,N} \rangle
= \frac{1}{\nu} \sum\limits_{m=1}^K \biggl(\frac{1}{b_m} + \sum\limits_{j=m+1}^K \frac{(\koff/\nu)^{j-m}}{b_j} 
\prod\limits_{i=m}^{j-1} \frac{d_i}{b_i}\biggr),
\end{equation}
with $b_m = N-K+m$ and $d_m = K-m$.  Lawley and Madrid proved that
$\bar{\N}(t)$ and $\bar{\T}_{K,N}$ are universal bounds to $\N(t)$ and
$\T_{K,N}$:
\begin{equation}
\N(t) \geq \bar{\N}(t) ~~\textrm{for all}~t, \qquad \T_{K,N} \leq \bar{T}_{K,N} ~~ \textrm{for all}~K ,
\end{equation}
which actually means that
\begin{eqnarray}
\P\{\N(t) > K\} &\geq& \P\{\bar{\N}(t) > K\} , \\
\P\{\T_{K,N} > t\} &\leq& \P\{ \bar{\T}_{K,N} > t\} .
\end{eqnarray}
Moreover, when the target region is getting smaller and smaller, these
bounds become more and more accurate.

The Lawley-Madrid approximation (LMA) opens a way to investigate in
detail the role of reversible binding onto the statistics of
sophisticated biochemical processes involving the arrival of several
molecules onto the target region.  The prominent example is the
signalling process between neurons when the fusion of a
neurotransmitters vesicle with the presynaptic bouton membrane is
triggered by the arrival of five calcium ions onto the sensor protein
\cite{Alberts,Berridge03,Eggermann12,Nakamura15,Dittrich13,Guerrier16,Reva21}.
It was recently shown by extensive simulations that unbinding events
considerably affect the fusion probability \cite{Reva21}.

In spite of numerous advantages of the LMA, it relies on a rough
assumption that both first-binding and rebinding times can be modeled
by an exponential random variable.  However, the probability density
of the rebinding time is in general more sophisticated; for instance,
in the case of a spherical target, it diverges at short times as $H(t)
\propto t^{-1/2}$ (see Appendix \ref{sec:Aspheres}), in sharp contrast
to the assumed exponential density $\nu e^{-\nu t}$ that behaves as
$\nu + O(t)$ as $t\to 0$.  This observation suggests that the LMA does
not capture correctly the short-time behavior that can be relevant for
some applications.  In this paper, we undertake a systematic study of
the problem of impatient particles in the case $K = N$.  Using the
renewal approach, we derive the exact solution for this problem.  We
deduce the short-time and long-time behavior of this solution and
compare it with the LMA predictions, as well as to Monte Carlo
simulations.  The short-time asymptotic analysis is also extended to
arbitrary $K$.  We show that the LMA captures the qualitative behavior
of the reaction time distribution at moderate and long times.
However, the LMA overestimates the mean reaction time and the decay
time, and fails at short times.  From the practical point of view, the
LMA can thus be used for qualitative estimations but further
improvements are necessary for getting more accurate results.

The paper is organized as follows.  In Sec. \ref{sec:main}, we start
with the mathematical model of impatient particles, derive the exact
form of the probability density of the reaction time $\T_{N,N}$, and
analyze its short-time and long-time asymptotic behavior.  We also
obtain the mean reaction time $\langle \T_{N,N}\rangle$ and the decay
time $T_N$ of the exponential decrease of the probability density at
long times.  We illustrate the obtained results for a relevant example
of restricted diffusion towards a spherical target.  Section
\ref{sec:discussion} is devoted to a systematic comparison of the
exact solution with two reference solutions: the irreversible binding
case and the LMA.  In Sec. \ref{sec:conclusions}, we summarize our
findings, discuss eventual applications, and provide final remarks and
perspectives.  Appendices contain technical details of the asymptotic
analysis (Sec. \ref{sec:asymptotics}), summary of formulas for
restricted diffusion between two concentric spheres
(Sec. \ref{sec:Aspheres}), numerical implementation of the exact
solution (Sec. \ref{sec:computation}) and description of Monte Carlo
simulations (Sec. \ref{sec:MC}).

\section{Main results}
\label{sec:main}

\subsection{Mathematical model of impatient particles}

We consider $N$ independent indistinguishable point-like particles
diffusing with diffusion coefficient $D$ inside a bounded Euclidean
domain $\Omega \subset \R^d$.  The boundary $\pa$ of $\Omega$ is
reflecting everywhere, except for a partially reactive target region
$\A$.  After hitting the target, a particle can bind to it with some
probability controlled by the reactivity $\kappa$
\cite{Collins49,Sano79,Shoup82,Zwanzig90,Sapoval94,Filoche99,Benichou00,Grebenkov03,Berezhkovskii04,Grebenkov06,Grebenkov06a,Reingruber09,Lawley15,Grebenkov17b,Bernoff18b,Grebenkov19b,Grebenkov20}.
This binding is reversible, i.e., the bound particle stays on the
target for an independent random waiting time distributed according to
the exponential law with the rate $\koff$.  After unbinding from the
target, the particle resumes its bulk diffusion from a random
uniformly distributed location on the target boundary, until the next
time it binds to the target.  We are interested in computing the
probability density $H_N(t|\x_0) = H_{N,N}(t|\x_0)$ of the first
instance $\T_N = \T_{N,N}$ when all of the $N$ particles are bound to
the target (i.e., when the process $\N(t)$ crosses the level $N$ for
the first time).  As the arrival of $N$ particles to the target is
supposed to trigger some reaction event, the first-crossing time
$\T_N$ is called the {\it reaction time.}  For simplicity, we assume
that all the particles are initially free (not bound to the target)
and start from the same initial position $\x_0 \in \Omega$.  These
starting assumptions can be easily relaxed.

\subsection{Exact solution}

To proceed, we introduce the probability $\Pr_t(n|m)$ that starting
from $m$ particles bound to the target at time $0$, there are $n$
particles bound to the target at time $t$.  This probability is hard
to compute in general due to unbinding events.  However, there are two
particular cases for which $\Pr_t(n|m)$ can be expressed in terms of a
single particle dynamics.  Let $P(t|\x_0)$ denote the occupancy
probability that an initially free particle that started from a point
$\x_0$ is bound to the target at time $t$.  Since all particles are
independent, the probability of finding $m$ bound particles on the
target at time $t$ is
\begin{equation}  \label{eq:Pr_m0}
    \Pr_t(m|0) = {N\choose m} [P(t|\x_0)]^m \big[1-P(t|\x_0)\big]^{N-m} ,
\end{equation}
where we have chosen the initial condition that all particles are
initially free.  Similarly, if $Q(t)$ denotes the probability that,
starting from the bound state at time $0$, the particle is bound to
the target at time $t$, then
\begin{equation}
    \Pr_t(m|N) = {N\choose m} [Q(t)]^m \big[1-Q(t)\big]^{N-m} .
\end{equation}
The probability density of the reaction time $\T_N$ can then be
obtained from a standard renewal equation:
\begin{equation}  \label{deconv}
   \Pr_t(N|0)= \int_{0}^{t} dt' \, H_{N}(t') \, \Pr_{t-t'}(N|N).
\end{equation}
Switching to Laplace space allows us to get
\begin{equation} \label{central1}
\tilde{H}_{N}(p|\x_0) =  \frac{\L \{ [P(t|\x_0)]^N \}}{\L \{[Q(t)]^N \}} \,,
\end{equation}
where both $\L$ and tilde denote the Laplace transform, e.g.
\begin{equation*}
\tilde{f}(p) = \L[f(t)] = \int\limits_0^\infty dt \, e^{-pt} \, f(t) .
\end{equation*}
The inversion of the Laplace transform gives the probability density
in time domain:
\begin{equation}  \label{central}
H_N(t|\x_0) = \L^{-1}  \biggr\{ \frac{\L \{ [P(t|\x_0)]^N \}}{\L \{[Q(t)]^N \}} \biggr\} .
\end{equation}

The last step consists in relating the probabilities $P(t|\x_0)$ and
$Q(t)$ to the first-passage time statistics of a single particle.
This can be done in a standard way by summing contributions according
to the number of unbinding events (see, e.g., \cite{Reva21}).  For
instance, one finds
\begin{equation*}
Q(t) = \Psi(t) + \int\limits_0^t dt_1 \, \psi(t_1) \int\limits_{t_1}^t dt'_1 \, H(t'_1-t_1) \, \Psi(t-t'_1) + \ldots ,
\end{equation*}
where $\Psi(t) = e^{-\koff t}$ is the probability of staying in the
bound state up to time $t$, $\psi(t) = - d\Psi(t)/dt = \koff e^{-\koff
t}$ is the probability density of the associated waiting time, and
$H(t)$ is the probability density of the rebinding time.  The first
term in the above equation is the contribution without unbinding.  In
the second term, the particle unbinds at time $t_1$, diffuses in the
bulk until the next rebinding at time $t'_1$, and remains bound until
time $t$.  The third and next terms correspond to 2, 3, etc. unbinding
events.  In Laplace domain, one simply gets
\begin{eqnarray}  \nonumber
\tilde{Q}(p) &=& \tilde{\Psi}(p) + \tilde{\psi}(p) \tilde{H}(p) \tilde{\Psi}(p) + \ldots  
= \frac{\tilde{\Psi}(p)}{1 - \tilde{\psi}(p) \tilde{H}(p)} \\  \label{eq:Qtilde}
&=& \frac{1}{p + \koff (1-\tilde{H}(p))} \,.
\end{eqnarray}
In turn, the occupancy probability $P(t|\x_0)$ includes an additional
step of the first-passage to the target that yields:
\begin{eqnarray}  \label{eq:Ptilde}
\tilde{P}(p|\x_0) &=& \tilde{H}(p|\x_0) \, \tilde{Q}(p) \\   \label{eq:Ptilde_bis}
&=& \frac{\tilde{H}(p|\x_0)}{p + \koff (1-\tilde{H}(p))} \,,
\end{eqnarray}
where $\tilde{H}(p|\x_0)$ is the Laplace transform of the probability
density $H(t|\x_0)$ of the first-passage time to the target when the
particle started from a point $\x_0$.  
Note that as the particle is released after unbinding from a uniformly
distributed point on the target boundary $\A$, one also gets
\begin{equation}
H(t) = \frac{1}{|\A|} \int\limits_{\A} d\x_0 \, H(t|\x_0)  ,
\end{equation}
where $|\A|$ is the Lebesgue measure of $\A$ (e.g., the area of $\A$
in the three-dimensional case).  In this way, both probabilities
$P(t|\x_0)$ and $Q(t)$ are expressed in terms of the first-passage
time probability density $H(t|\x_0)$ for a single particle.
In the case $N = 1$, comparison of Eqs. (\ref{central1},
\ref{eq:Ptilde}) yields immediately that $H_1(t|\x_0) = H(t|\x_0)$, as
expected.

\subsection{Spectral decompositions}

As we deal with restricted diffusion in a bounded domain, the
probabilities $P(t|\x_0)$ and $Q(t)$ can be formally deduced from
their Laplace transforms by applying the residue theorem.  Let
$\{p_n\}$ be the poles of $\tilde{P}(p|\x_0)$ that lie on the {\it
negative} real axis: $0 = p_0 > p_1 \geq p_2 \geq \searrow - \infty$.
According to Eq. (\ref{eq:Ptilde_bis}), these poles satisfy the
equation:
\begin{equation}  \label{eq:poles}
p_n + \koff (1 - \tilde{H}(p_n)) = 0 .
\end{equation}
Note that since the poles of $\tilde{H}(p|\x_0)$ and $\tilde{H}(p)$
are the same, they cancel each other in Eq. (\ref{eq:Ptilde_bis}) and
thus are not included in the set of poles of $\tilde{P}(p|\x_0)$.
If all the poles are simple, the inverse Laplace transform yields
\begin{equation}  \label{eq:P_spectral}
P(t|\x_0) = P_\infty + \sum\limits_{n=1}^\infty v_n(\x_0) \, e^{p_n t} ,
\end{equation}
where $v_n(\x_0)$ is the residue of $\tilde{P}(p|\x_0)$ evaluated at
the pole $p_n$.  The steady-state limit $P_\infty$ corresponds to the
pole at $0$, which can be obtained by using the Taylor expansion
\begin{equation}
\tilde{H}(p|\x_0) = \langle e^{-p\tau}\rangle_{\x_0} = 1 - p \langle \tau\rangle_{\x_0} + O(p^2),
\end{equation}
where $\langle \tau\rangle_{\x_0}$ is the mean FPT to the target for a
single particle started from $\x_0$.  Similarly, $\tilde{H}(p) = 1 - p
\langle \tau\rangle + O(p^2)$, where $\langle \tau\rangle$ is the mean
rebinding time: 
\begin{equation}
\langle \tau \rangle = \frac{1}{|\A|} \int\limits_{\A} d\x_0 \, \langle \tau\rangle_{\x_0} .
\end{equation}
As a consequence, Eq. (\ref{eq:Ptilde_bis}) implies that
\begin{equation}  \label{eq:Pinf}
P_\infty = \frac{1}{1 + \koff \langle \tau\rangle} \,.  
\end{equation}

The mean rebinding time $\langle \tau \rangle$ can be found explicitly
by writing the boundary value problem for the mean FPT:
\begin{equation}
\left\{  \begin{array}{r l l}
D \Delta \langle \tau\rangle_{\x_0} & = - 1 & (\x_0 \in \Omega), \\  
- D \partial_n \langle \tau\rangle_{\x_0} & =  \kappa \I_{\A}(\x_0) \langle \tau\rangle_{\x_0}   & (\x_0\in \pa), \\ \end{array} \right.
\end{equation}
where $\I_{\A}(\x_0)$ is the indicator function of $\A$:
$\I_{\A}(\x_0) = 1$ if $\x_0 \in \A$, and $0$ otherwise.  Integrating
the first relation over $\x_0 \in \Omega$ and applying the Green's
formula, one gets
\begin{eqnarray*}
-|\Omega| &=& \int\limits_{\Omega} d\x_0 \, D \Delta \langle \tau\rangle_{\x_0} = 
\int\limits_{\pa} d\x_0 \, D \partial_n \langle \tau\rangle_{\x_0} \\
&=& - \int\limits_{\A} d\x_0 \, \kappa \langle \tau\rangle_{\x_0} = - \kappa |\A| \langle \tau\rangle ,
\end{eqnarray*}
from which
\begin{equation}  \label{eq:mean_rebinding}
\langle \tau\rangle = \frac{|\Omega|}{\kappa |\A|}  \,,
\end{equation}
where $|\Omega|$ is the volume of the domain.  For a spherical target
of radius $\rho$, the reactivity can be expressed in terms of the
forward constant $\kon = \kappa (4\pi \rho^2 N_A)$ (with $N_A
\approx 6.02 \cdot 10^{23}~\rm{mol}^{-1}$ being the Avogadro number)
\cite{Shoup82,Lauffenburger} so that the mean rebinding time also reads
as $\langle \tau \rangle = N_A |\Omega|/\kon$.  Defining the
dimensionless quantity
\begin{equation}  \label{eq:eta}
\eta = \koff \langle \tau \rangle = \frac{\koff |\Omega|}{\kappa |\A|} = \frac{\koff |\Omega| N_A}{\kon} \,,
\end{equation}
we simply get $P_\infty = 1/(1 + \eta)$.

In general, the poles are not necessarily simple.  In particular, if
the unbinding rate $\koff$ is such that $\tilde{H}(-\koff) = 0$, then
$-\koff$ is the pole of $\tilde{P}(p|\x_0)$ of higher order than $1$.
For instance, if $-\koff$ is the pole of order $2$, the corresponding
term in the spectral expansion (\ref{eq:P_spectral}) is of the form $t
e^{-\koff t}$.  As the set of zeros of the function $\tilde{H}(p)$ is
discrete, we will ignore such specific values of the unbinding rate
$\koff$.

Introducing
\begin{equation}  \label{eq:Ptilde_Qtilde}
\tilde{P}(p) = \frac{1}{|\A|} \int\limits_{\A} d\x_0 \, \tilde{P}(p|\x_0) = \tilde{H}(p) \, \tilde{Q}(p),
\end{equation}
we can express $\tilde{Q}(p)$ from Eq. (\ref{eq:Qtilde}) as
\begin{equation}
    \tilde{Q}(p) = \frac{1}{p+\koff} + \frac{\koff}{p+\koff} \tilde{P}(p),
\end{equation}
which in time domain reads
\begin{equation}  \label{eq:Qt}
    Q(t) = e^{-\koff t} + \koff \int\limits_0^t dt' \, e^{-\koff t'} \,P(t-t')  .
\end{equation}
This relation implies that $Q(t)$ monotonously decreases from $Q(0) =
1$ to $Q(\infty) = P_\infty$ (see Appendix \ref{sec:Qt}).
Substituting Eq. (\ref{eq:P_spectral}) into this relation, we get
\begin{equation}  
Q(t) = P_\infty + Q_0  e^{-\koff t} + \sum\limits_{n=1}^\infty q_n e^{p_n t} ,
\end{equation}
where
\begin{eqnarray*}
Q_0 &=&  1-P_\infty - \sum\limits_{n=1}^\infty q_n \,, \\
q_n &=&  \frac{v_n}{1 + p_n/\koff}  \quad (n=1,2,\ldots), \\
v_n &=& \frac{1}{|\A|} \int\limits_{\A} d\x_0 v_n(\x_0) ,
\end{eqnarray*}
and we assumed that $-\koff$ is not the pole.  On one hand, evaluating
$\tilde{P}(p)$ at $p = -\koff$, one finds
\begin{equation*}
\tilde{P}(-\koff) = \frac{P_\infty}{-\koff} - \sum\limits_{n=1}^\infty \frac{v_n}{\koff + p_n}  \,.
\end{equation*}
On the other hand, Eq. (\ref{eq:Ptilde_bis}) implies
\begin{equation*}
\tilde{P}(-\koff) = \frac{\tilde{H}(-\koff)}{-\koff + \koff(1-\tilde{H}(-\koff))} = - \frac{1}{\koff} \,,
\end{equation*}
yielding $Q_0 \equiv 0$, and thus
\begin{equation}   \label{eq:Qt_spectral}
Q(t) = P_\infty + \sum\limits_{n=1}^\infty q_n e^{p_n t} .
\end{equation}

In summary, Eqs. (\ref{central}, \ref{eq:P_spectral},
\ref{eq:Qt_spectral}) fully determine the exact form of the
probability density $H_N(t|\x_0)$ in terms of the first-passage time
statistics $H(t|\x_0)$ of a single particle.  Even if $H(t|\x_0)$ is
known explicitly (see an example in Appendix \ref{sec:Aspheres}), a
numerical implementation of this exact solution remains challenging
because it involves: finding zeros $\{p_n\}$ of Eq. (\ref{eq:poles}),
evaluation of the residues at these poles, computation of spectral
expansions (\ref{eq:P_spectral}, \ref{eq:Qt_spectral}) and finally the
inverse Laplace transform in Eq. (\ref{central}).  The practical
details of this computation are discussed in Appendix
\ref{sec:computation}.  At the same time, our exact solution opens a
way to investigate the asymptotic behavior of the exact probability
density $H_N(t|\x_0)$ in a rather general setting.  Before turning to
this analysis, we discuss the mean reaction time.

\subsection{Mean reaction time}
\label{sec:mean}

The relation (\ref{central}) allows one to access the moments of the
reaction time:
\begin{equation}
    \langle \T_N^k \rangle = (-1)^k \lim_{p\to 0} \frac{\partial^k}{\partial p^k}  \,\frac{\L\{ [P(t|\x_0)]^N \}}{\L\{ [Q(t)]^N \}} \,.
\end{equation}
In particular, the mean reaction time is
\begin{eqnarray*}
\langle \T_N \rangle &=& \lim\limits_{p\to 0} \biggl(\frac{\L\{ t [P(t|\x_0)]^N\}}{\L\{ [Q(t)]^N\}} \\
&-& \frac{\L\{ [P(t|\x_0)]^N\} \, \L\{ t[Q(t)]^N\}} {(\L\{ [Q(t)]^N\})^2} \biggr).
\end{eqnarray*}
As both $P(t|\x_0)$ and $Q(t)$ tend to $P_\infty$ in the long-time
limit, setting $p = 0$ in the above Laplace transforms would yield
divergence.  To overcome this issue, one can add and subtract the term
$P_\infty^N$ to each Laplace transform, e.g.,
\begin{eqnarray*}
\L\{ t [P(t|\x_0)]^N\} &=& \int\limits_0^\infty dt \, t\, e^{-pt} \bigl([P(t|\x_0)]^N - P_\infty^N + P_\infty^N\bigr) \\
&=& P_\infty^N/p^2 + a_1 + o(1)  \qquad (p\to 0),
\end{eqnarray*}
where 
\begin{equation}
a_k = \int\limits_0^\infty dt \, t^k \, \bigl([P(t|\x_0)]^N - P_\infty^N\bigr).
\end{equation}
Introducing also
\begin{equation}
b_k = \int\limits_0^\infty dt \, t^k \, \bigl([Q(t)]^N - P_\infty^N\bigr) ,
\end{equation}
we compute the above limit as
\begin{equation}  \label{eq:TN_mean}
\langle \T_N \rangle = \frac{b_0 - a_0}{P_\infty^N} = \int\limits_0^\infty dt \, \frac{[Q(t)]^N - [P(t|\x_0)]^N}{P_\infty^N} \,.
\end{equation}
Higher-order moments of $\T_{N}$ can be expressed in a similar way.
For $N = 1$, this relation implies that $\langle \T_1\rangle = \langle
\tau\rangle_{\x_0}$, as expected.  Equation (\ref{eq:TN_mean}) is a
generalization of the expression for the mean slowest FPT governed by
the probability density in Eq. (\ref{eq:HKN}):
\begin{equation}  \label{eq:Tmean_irrev}
\langle \Ti_{N,N} \rangle = \int\limits_0^\infty dt \bigl(1 - [1-S(t)]^N\bigr).
\end{equation}
In fact, if there is no unbinding ($\koff = 0$), one gets $Q(t) = 1$,
$P_\infty = 1$, and $P(t|\x_0) = 1 - S(t|\x_0)$.  In Appendix
\ref{sec:Tmean_irrev}, we derive the large-$N$ asymptotic behavior of
this mean time: 
\begin{equation}
\langle \Ti_{N,N} \rangle \propto \langle \tau \rangle \ln N  \qquad (N\to \infty).
\end{equation}
The unbinding mechanism drastically changes this asymptotic behavior
into
\begin{equation}  \label{eq:Tmean_asympt}
\langle \T_{N} \rangle \propto \frac{(1 + \koff \langle \tau \rangle)^N}{\koff N}  \qquad (N\to \infty),
\end{equation}
i.e., a very slow logarithmic increase turns into exponential growth
controlled by the unbinding rate $\koff$ (see Appendix
\ref{sec:ATmean}).  As a consequence, when many particles are needed
to trigger the reaction, even a small unbinding rate can considerably
alter predictions of the irreversible setting.  We emphasize however,
that Eq. (\ref{eq:Tmean_asympt}) captures only the large-$N$
asymptotic behavior and is not applicable at small $N$.  In
particular, an non-monotonous dependence of the right-hand side of
Eq. (\ref{eq:Tmean_asympt}) on $\koff$ and $N$ is not reproduced for
the mean reaction time (see further discussion in Appendix
\ref{sec:ATmean}).

\subsection{Long-time behavior}
\label{sec:long}

The probability density $H_N(t|\x_0)$ can be formally obtained via the
inverse Laplace transform in Eq. (\ref{central1}) by finding the poles
$p_{n,N}$ of the function $\tilde{H}_N(p|\x_0)$ in the complex plane
and applying the residue theorem.  This is a difficult task, even
numerically, especially for large $N$.  We focus therefore on the pole
$p_{1,N}$ with the smallest absolute value that determines the decay
time $T_N = 1/|p_{1,N}|$ of the probability density at long times:
\begin{equation}  \label{eq:HN_long}
H_N(t|\x_0) \propto e^{- t/T_N} \qquad (t\to\infty). 
\end{equation}

As $P(t|\x_0)$ admits the spectral decomposition (\ref{eq:P_spectral})
with the poles $p_n$, the poles of the numerator $\L\{[P(t|\x_0)]^N\}$
of Eq. (\ref{central1}) are obtained as all linear combinations of the
form $p_{n_1} + p_{n_2} + \ldots + p_{n_N}$.  In particular, the pole
with the smallest absolute value is still $p_1$ (apart from the pole
at $0$).  The situation is more difficult for the denominator
$\L\{[Q(t)]^N\}$, for which we are looking not for its poles, but for
zeros.
Let us search for a zero of this function:
\begin{equation}
\L\{[Q(t)]^N\}(p) = \sum\limits_{n_1=0}^\infty \cdots \sum\limits_{n_N=0}^\infty 
\frac{q_{n_1} \cdots q_{n_N}}{p - p_{n_1} - \ldots - p_{n_N}} \,,
\end{equation}
where we included the pole at $0$ by setting $p_0 = 0$ and $q_0 =
P_\infty$.  As $p\to 0$, the leading term of this expression is
$q_0^N/p$, which can be separated from the other terms.  In the
leading-order approximation, one can set $p = 0$ in the remaining
terms:
\begin{equation}  \label{eq:auxil12}
0 = \L\{[Q(t)]^N\}(p_{1,N}) \approx \frac{q_0^N}{p_{1,N}} - \hspace*{-5mm} \sum\limits_{\substack{n_1=0, \ldots \\ \ldots, n_N=0 ,
\\ n_1+\ldots+n_N > 0}}^\infty \hspace*{-3mm}   \frac{q_{n_1} \cdots q_{n_N}}{p_{n_1} + \ldots + p_{n_N}} .
\end{equation}
The multiple sum, from which the term with $n_1 = n_2 = \ldots = n_N =
0$ was subtracted, can be expressed in terms of an integral, yielding
an approximation for the pole $p_{1,N}$:
\begin{equation}  \label{eq:p1_approx}
p_{1,N} \approx - P_\infty^N \biggl(\int\limits_0^\infty dt \bigl([Q(t)]^N - P_\infty^N\bigr) \biggr)^{-1} .
\end{equation}
As a consequence, the decay rate is
\begin{equation}  \label{eq:TN_decay}
T_N \approx P_\infty^{-N} \int\limits_0^\infty dt \bigl([Q(t)]^N - P_\infty^N\bigr) .
\end{equation}
Curiously, this expression is very similar to the expression
(\ref{eq:TN_mean}) for the mean reaction time.  

The accuracy of this approximation depends on various parameters such
as $\koff$ and $N$.  In fact, in order to get the sum in
Eq. (\ref{eq:auxil12}), we neglected $p_{1,N}$ under the assumption
that $|p_{1,N}|$ is much smaller than $|p_1|$.  As $N$ increases, the
reaction event occurs at longer times, i.e., the decay time increases,
and the approximation gets more accurate.  In turn, the case $N = 1$
is the worst for this approximation (see discussion in Appendix
\ref{sec:caseN1}).  Similarly, as $\koff$ increases, the particles
unbind more often, the decay time increases, yielding a more accurate
approximation.  Note that the approximation (\ref{eq:p1_approx}) can
be improved by accounting perturbatively for next-order corrections.

\subsection{Short-time behavior}
\label{sec:short}

At short times, the main contribution to the probability density of
the first-passage time comes from the particles that follow almost
``direct trajectories'' to the target
\cite{Godec16b,Basnayake18,Grebenkov18}.  As a consequence, the
short-time behavior is generally
\begin{equation}  \label{hrt_short}
H(t|\x_0) \approx C_{\x_0} \, t^\alpha \, e^{-\delta^2/(4Dt)}  \qquad (t\to 0),
\end{equation}
where $\delta$ is the distance between the starting point $\x_0$ and
the target region $\A$, $t^\alpha$ is a power-law correction, and
$C_{\x_0}$ is the prefactor depending on the starting point, the shape
of the domain, and the reactivity $\kappa$.  Note that $\delta$ is
either the Euclidean distance (i.e., the length of the shortest
interval connecting $\x_0$ and $\A$), or the geodesic distance along
the shortest curvilinear path from $\x_0$ and $\A$ that bypasses
eventual obstacles.  As a rigorous derivation of this relation is
beyond the scope of the paper, we use it as an assumption, under which
the following results are valid (see an example in Appendix
\ref{sec:spheres_short}).

When the starting point $\x_0$ lies on the target, $\delta = 0$ and
Eq. (\ref{hrt_short}) implies $H(t) \approx C t^\alpha$ with $C =
(1/|\A|) \int\nolimits_{\A} d\x_0 \, C_{\x_0}$, from which
$\tilde{H}(p) \approx C \Gamma(\alpha+1) p^{-1-\alpha}$ as
$p\to\infty$.  If $\alpha > -2$, the leading term in the denominator
of Eq. (\ref{eq:Ptilde_bis}) is $p$ that implies for any $\x_0$:
\begin{equation}
    \tilde{P}(p|\x_0) \approx \frac{\tilde{H}(p|\x_0)}{p} \,,
\end{equation}
which in time domain gives 
\begin{equation}  \label{prt_short0}
    P(t|\x_0) \approx \int_{0}^{t} dt' \, H(t'|\x_0) = 1 - S(t|\x_0).
\end{equation}

When $\x_0 \notin \A$, the integral of Eq.~\eqref{hrt_short} yields in
the lowest order
\begin{equation}      \label{prt_short}
    P(t|\x_0) \approx \frac{4D C_{\x_0}}{\delta^2}  \, t^{\alpha+2} \, e^{-\delta^2/(4Dt)} \,,
\end{equation}
from which Eq. (\ref{eq:Pr_m0}) implies
\begin{equation}
\Pr_t(N|0) \approx \left(\frac{4D C_{\x_0}}{\delta^2} \right)^N \, t^{N(\alpha+2)} \,e^{-N \delta^2/(4Dt)} .
\end{equation}
At short times, one has $Q(t)\approx 1$, and thus $\Pr_t(N|N)\approx
1$, so that
\begin{equation}
    \tilde{H}_N(p|\x_0) \approx  p\mathcal{L}\big\{\Pr_t(N|0)\big\} \,,
\end{equation}
which in turn gives us the short-time behavior
\begin{eqnarray}  \nonumber
&&  H_N(t|\x_0) \approx \frac{\partial{\Pr_t(N|0)}}{\partial{t}} \\   \label{eq:HN_short}
&& \approx  \frac{N \delta^2}{4D}  \left(\frac{4D C_{\x_0}}{\delta^2} \right)^N \, t^{N(\alpha+2)-2} \, e^{-N \delta^2/(4Dt)} .    
\end{eqnarray}
This leading-order asymptotic relation can be improved by computing
the next-order term in the integral of Eq. (\ref{hrt_short}) that
yields the correction $O(t)$ to Eq. (\ref{eq:HN_short}), which is
still independent of the unbinding rate $\koff$.  In turn, $\koff$
appears in the correction $O(t^2)$ by using $Q(t) \approx 1 - \koff t$
instead of $Q(t) \approx 1$ in the above derivation.  The integral of
this expression yields, in the leading order:
\begin{equation}  \label{eq:SN_short}
S_{N}(t|\x_0) \approx 1 - \left(\frac{4D C_{\x_0}}{\delta^2} \right)^N  t^{N(\alpha+2)} \, e^{- N \delta^2/(4Dt)} .
\end{equation}

It is easy to see that for $N=1$, the leading term of the short-time
behavior in Eq. \eqref{hrt_short} is recovered.  We note that the
short-time behavior of the reaction time density $H_N(t|\x_0)$ is
identical to that of the probability density $H^0_{N,N}(t)$ of the
first-passage time $\Ti_{N,N}$.  This result is independent of the
unbinding rate $\koff$ because the probability of an unbinding event
is small at times $t \ll \koff^{-1}$.  As a consequence, for any $K$,
one can approximate the short-time behavior of $\T_{K,N}$ by that of
$\Ti_{K,N}$, for which the probability density is given explicitly by
Eq. (\ref{eq:HKN}).  Substituting here the short-time asymptotic
relations (\ref{hrt_short}, \ref{prt_short0}, \ref{prt_short}), we get
then
\begin{eqnarray}  \label{eq:HKN_short}
&& H_{K,N}(t) \approx H^0_{K,N}(t) \\   \nonumber
&& \approx {N \choose K} \frac{K \delta^2}{4D}   \left(\frac{4D C_{\x_0}}{\delta^2} \right)^K  t^{K(\alpha+2) -2} \, e^{- K \delta^2/(4Dt)} ,
\end{eqnarray}
which generalizes Eq. (\ref{eq:HN_short}).

\section{Discussion}
\label{sec:discussion}

In order to illustrate our general results, we consider a relevant
example of restricted diffusion inside a reflecting sphere of radius
$R$ towards a partially reactive spherical target of radius $\rho$
located at the origin.  This geometrical setting is a simplified model
of passive diffusion inside the cytoplasm towards the nucleus.  It was
also employed to model diffusion of calcium ions inside a presynaptic
bouton towards a calcium-sensing protein \cite{Reva21}.  The
distribution of the first-passage time of a single particle was
investigated in \cite{Grebenkov18}.  In the presence of unbinding
events, the exact spectral decompositions for both probabilities
$P(t|\x_0)$ and $Q(t)$ were derived in \cite{Reva21}.  Appendix
\ref{sec:Aspheres} summarizes former results needed for studying
the problem of impatient particles.

The numerical method for evaluating the probability density
$H_N(t|\x_0)$ in Eq. (\ref{central}) is described in Appendix
\ref{sec:computation}.  To validate the accuracy of this exact
solution, we also performed Monte Carlo simulations, as described in
Appendix \ref{sec:MC}.  In the following, we set $\rho = 1$ and $D =
1$ to fix the units of length and time.  The radius of outer
reflecting sphere is set as $R = 10$ so that the target is relatively
small.  All the particles start from a fixed point $\x_0$ such that
either $|\x_0| = 5$ (relatively far from the target), or $|\x_0| = 2$
(relatively close to the target).  To analyze the effect of unbinding
events, we fix the reactivity $\kappa = 1$ (and thus the forward
constant $\kon$) and vary the unbinding rate $\koff$.  For $\kappa =
1$, the mean rebinding time $\langle \tau \rangle$ in
Eq. (\ref{eq:mean_rebinding}) is equal to $333$.  Setting $\koff =
0.003$ or $\koff = 0.03$, we can thus examine two settings of moderate
($\eta = 1$) and strong ($\eta = 10$) unbinding kinetics,
respectively.  We will compare our exact solution in
Eq. (\ref{central}) with Monte Carlo simulations, the LMA, the
irreversible binding solution, and the short-time asymptotic relation.

Table \ref{tab:mean} presents the mean reaction time for $N = 2$ and
$N = 3$ with two unbinding rates $\koff$, showing an excellent
agreement between Eq. (\ref{eq:TN_mean}) and Monte Carlo simulations.
In turn, the LMA overestimates the mean reaction time, the largest
deviation corresponding to stronger unbinding $\koff$ and larger $N$.
In addition, Table \ref{tab:mean} presents the decay time in the same
setting.  Expectedly, our approximation (\ref{eq:TN_decay}) is least
accurate for $N = 2$ and the small unbinding rate $\koff = 0.003$ (see
Sec. \ref{sec:long}).  At $N = 3$, the agreement is better.  Moreover,
for faster unbinding with $\koff = 0.03$, the approximation
(\ref{eq:TN_decay}) is in excellent agreement with the exact values
for both $N = 2$ and $N = 3$.  In contrast, the LMA predictions are
much less accurate.

\begin{table}
\begin{center}
\begin{tabular}{|c|c|c|c|c|c|c|c|}  \hline
\multicolumn{2}{|c|}{}  & \multicolumn{3}{c|}{Mean $\langle \T_{N}\rangle (\times 10^{3})$} 
& \multicolumn{3}{c|}{Decay time $T_N (\times 10^{3})$} \\ \hline 
$N$ & $\koff$ &  Theory  &  MC  & LMA  & Theory & Approx. & LMA \\  \hline
\multirow{2}{*}{$2$} 
   &  0.003  & 1.20 & 1.21 & 1.47 & 1.04 & 0.76 & 1.33 \\
   &  0.03   & 3.90 & 3.94 & 6.45 & 3.87 & 3.83 & 6.42 \\  \hline 
\multirow{2}{*}{$3$} 
   &  0.003  & 2.01 & 2.03 & 3.08 & 1.71 & 1.46 & 2.83 \\
   &  0.03   & 27.9 & 28.3 & 81.3 & 28.0 & 27.8 & 81.3 \\  \hline 
\end{tabular}
\end{center}
\caption{
Mean reaction time $\langle \T_N\rangle$ and the decay time $T_N$ for
restricted diffusion towards a spherical target of radius $\rho = 1$
and reactivity $\kappa = 1$, surrounded by a reflecting concentric
sphere of radius $R = 10$, for $N$ particles started from $|\x_0| = 5$
with $D = 1$ (see Appendix \ref{sec:Aspheres} for details).  Monte
Carlo (MC) values of $\langle \T_N\rangle$ were estimated from $10^6$
realizations (see Appendix \ref{sec:MC}), its theoretical values were
obtained by numerical integration of Eq. (\ref{eq:TN_mean}), while LMA
values were given by Eq. (\ref{eq:LMA_mean}).  Theoretical values of
the decay time $T_N$ were estimated by fitting
$S_N(t|\x_0)/H_N(t|\x_0)$ over a selected range of times, approximate
values were obtained by numerical integration of
Eq. (\ref{eq:TN_decay}), while the LMA values were deduced from the
inverse of the smallest eigenvalue of the matrix $-W^{(N)}$.  Note
that all times in the table should be multiplied by $10^3$.}
\label{tab:mean}
\end{table}

\begin{figure*}
\begin{center}
\includegraphics[width=80mm]{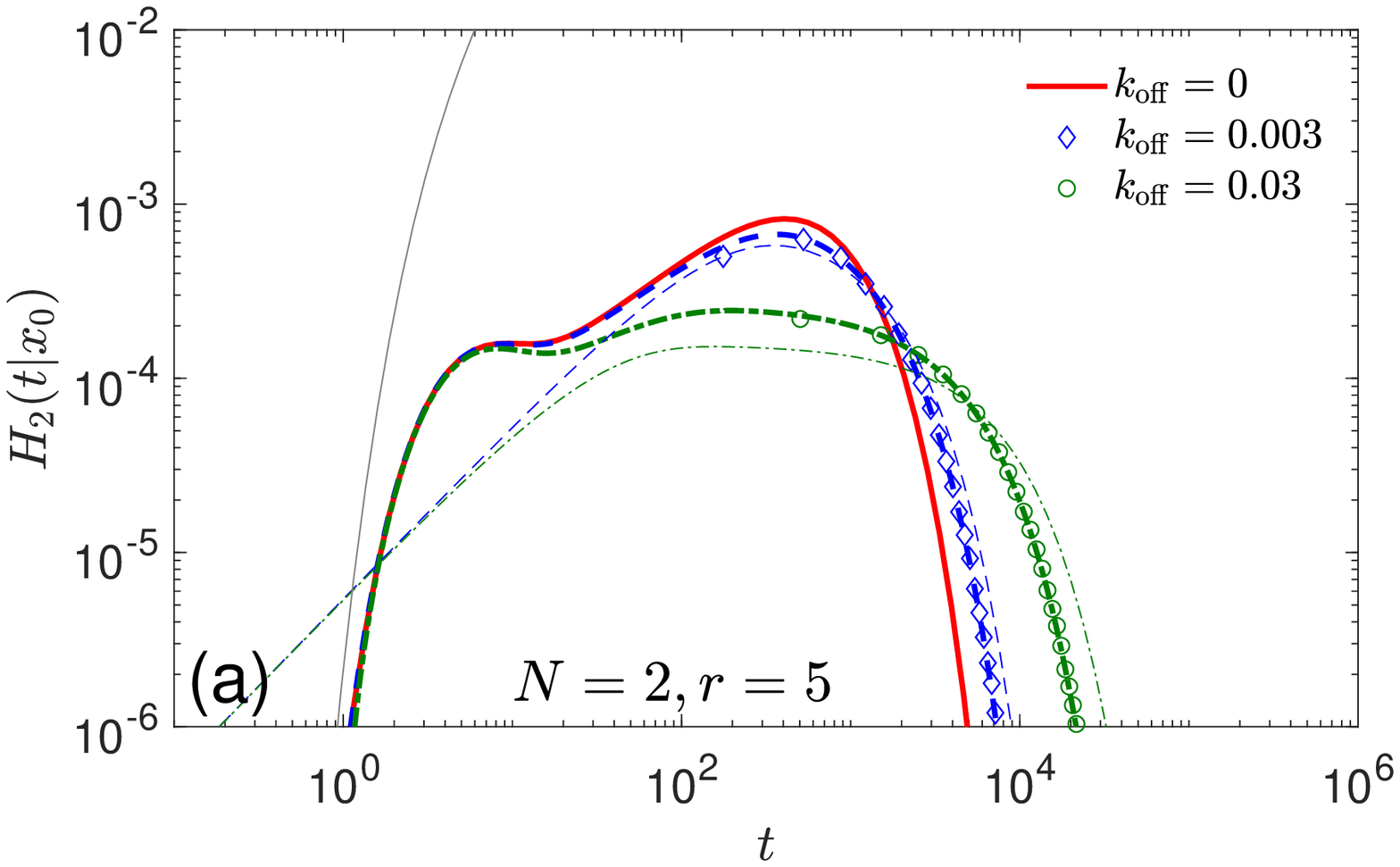} 
\includegraphics[width=80mm]{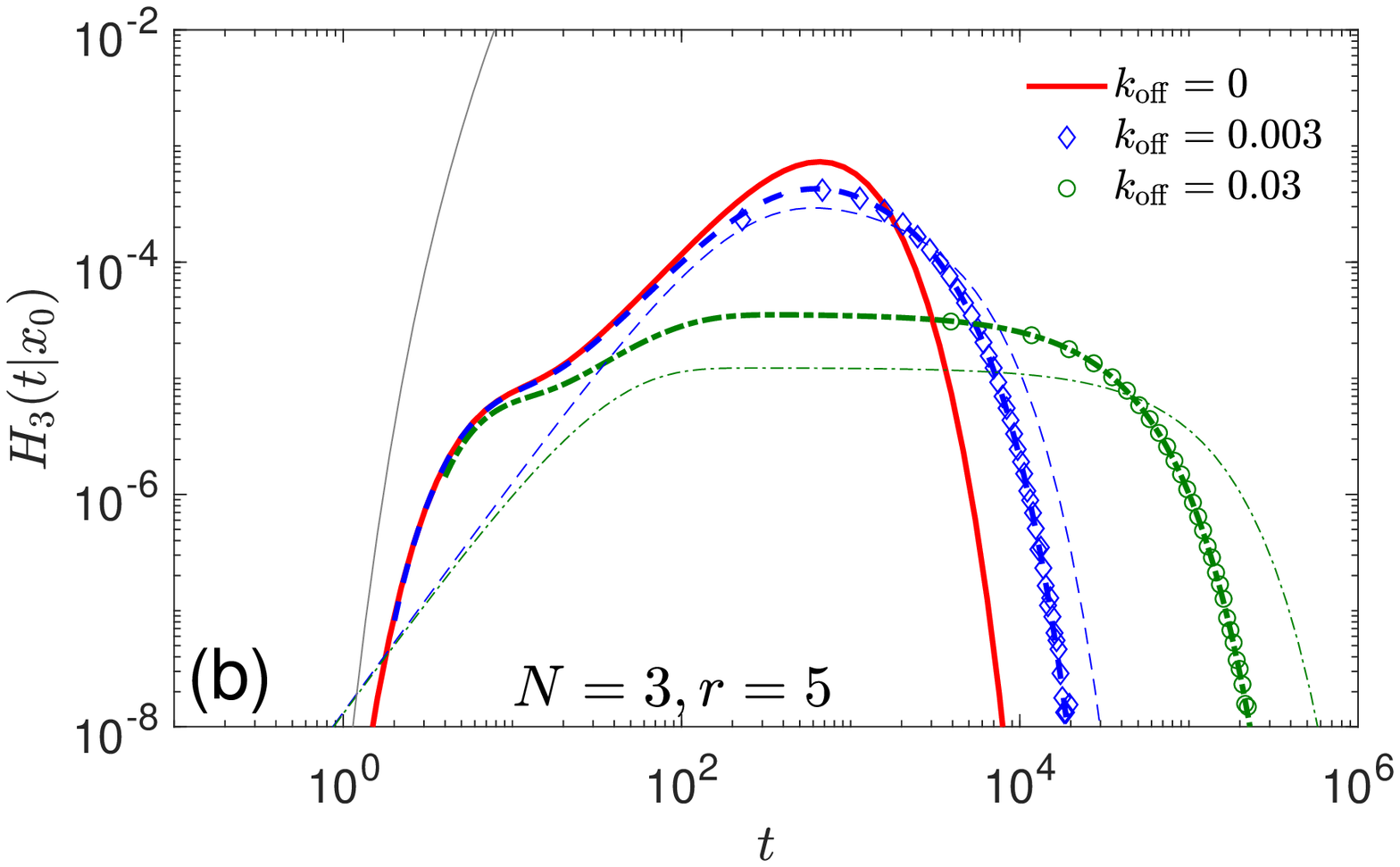} 
\includegraphics[width=80mm]{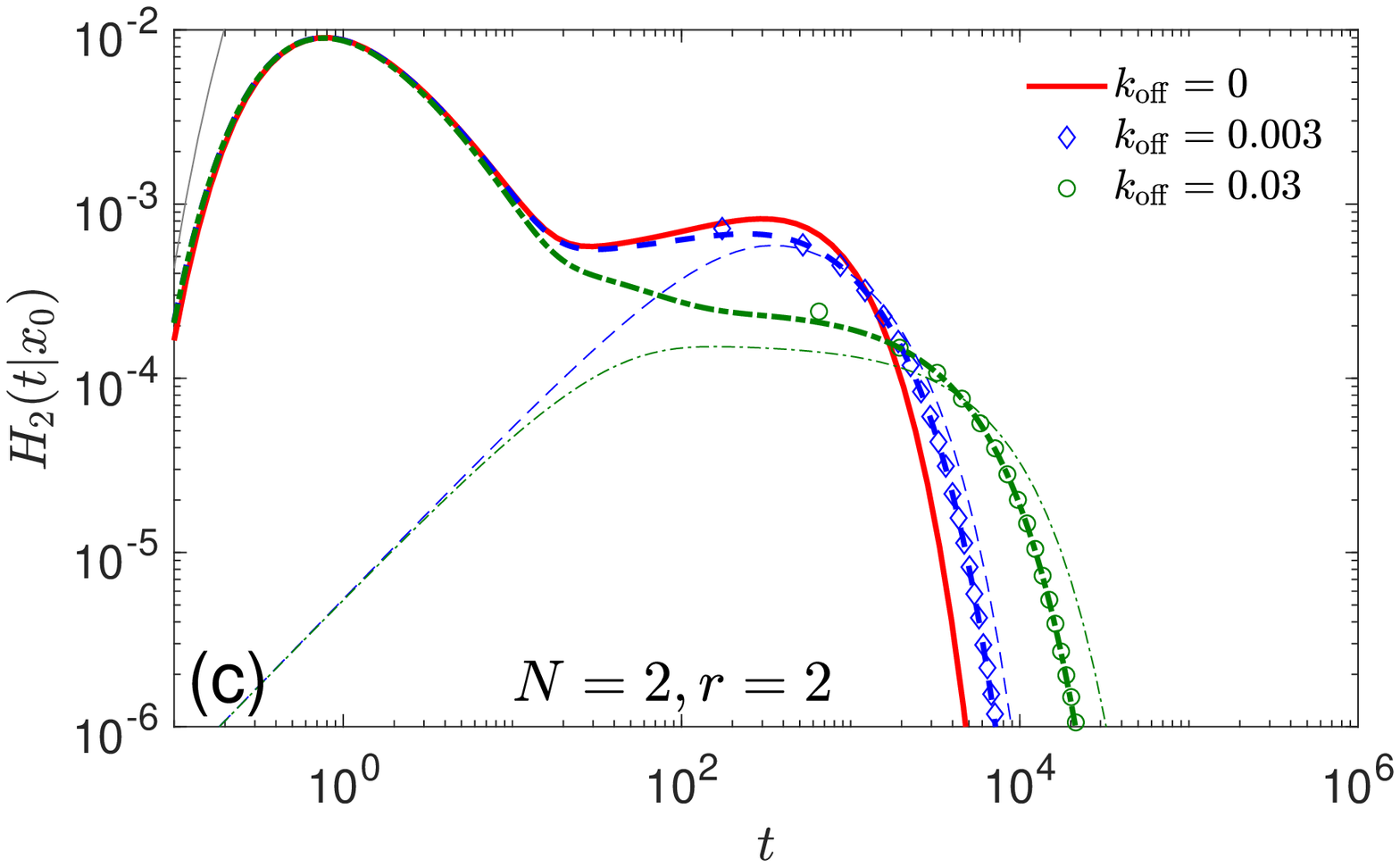} 
\includegraphics[width=80mm]{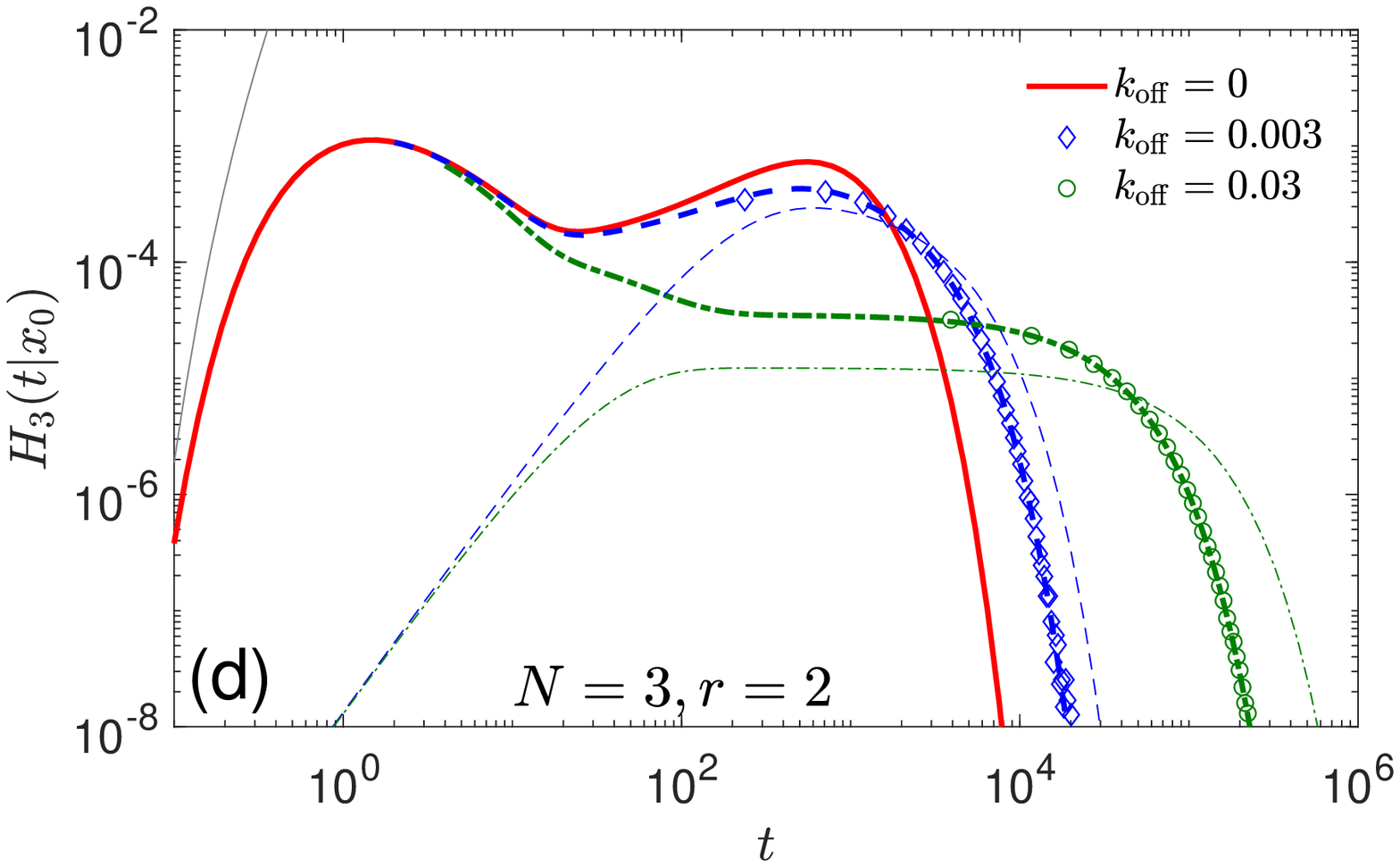} 
\end{center}
\caption{
Probability density $H_N(t|\x_0)$ of the reaction time $\T_N$ for
restricted diffusion between concentric spheres of radii $\rho = 1$
and $R = 10$, with $D = 1$, $\kappa = 1$, three values of $\koff$ (see
legend), two starting positions $|\x_0| = 5$ {\bf (a,b)} and $|\x_0| =
2$ {\bf (c,d)}, and two values of $N$: $N = 2$ {\bf (a,c)} and $N = 3$
{\bf (b,d)}.  Symbols show empirical histograms from Monte Carlo
simulations with $10^6$ particles.  Thick lines indicate our exact
solution (\ref{central}) evaluated numerically as described in
Appendix \ref{sec:computation}, whereas thin lines show the
Lawley-Madrid approximation (\ref{eq:HKN_LMA}).  Thin gray solid line
presents the short-time asymptotic behavior (\ref{eq:HN_short}). }
\label{fig:HN}
\end{figure*}

\subsection{Comparison with irreversible binding case}

First, we note that the limit of irreversible binding can be achieved
by setting either $\koff = 0$ or $\kappa = \infty$.  In fact, in the
latter case, any particle that unbinds from the target, immediately
rebinds and thus never leaves the target.  As a consequence, the
natural parameter characterizing the unbinding kinetics is the
dimensionless quantity $\eta$ defined by Eq. (\ref{eq:eta}).  When
$\eta$ is small, unbinding kinetics is usually considered as
irrelevant.  In the following, we consider the irreversible binding
limit by keeping $\kappa$ fixed and setting $\koff\to 0$.

For irreversible binding, the short-time behavior of the probability
density $H_{K,N}(t|\x_0)$ is given by Eq. (\ref{eq:HKN_short}).  In
turn, the long-time behavior follows from the spectral expansion of
the probability density $H(t|\x_0)$ of the first-passage time.  In
fact, as restricted diffusion occurs in a bounded domain, the
governing Laplace operator, $-\Delta$, has a discrete spectrum, i.e.,
a countable set of eigenvalues $0< \lambda_1 \leq \lambda_2 \leq
\lambda_3 \leq \ldots \nearrow \infty$ that are associated to
$L_2(\Omega)$-normalized eigenfunctions $\{u_n(\x)\}$ forming a
complete orthonormal basis in $L_2(\Omega)$ \cite{Grebenkov13}.  As a
consequence, the survival probability admits the standard spectral
decomposition \cite{Gardiner}
\begin{equation}  \label{eq:S_spectral}
S(t|\x_0) = \sum\limits_{n=1}^\infty c_n\,  u_n(\x_0) e^{-Dt\lambda_n}  , \quad c_n = \int\limits_\Omega d\x \, u_n(\x) ,
\end{equation}
from which Eq. (\ref{eq:HKN}) implies
\begin{equation}
H^0_{N,N}(t|\x_0)\approx N \, D\lambda_1\, c_1 u_1(\x_0) e^{- Dt\lambda_1}  \,.
\end{equation}
When the target is small, one has
\begin{equation}  \label{eq:lambda1_small}
\lambda_1 \approx \frac{\kappa |\A|}{D|\Omega|} = \frac{1}{D\langle \tau\rangle} \,,
\end{equation}
where we used Eq. (\ref{eq:mean_rebinding}) for the mean rebinding
time.  We get therefore
\begin{equation}
H^0_{N,N}(t|\x_0)\propto e^{-t/\langle \tau\rangle}  \qquad (t\to \infty).
\end{equation}
One sees that the decay time here, $\langle \tau \rangle$, does not
depend on $N$, in sharp contrast to the exponential growth of $T_N$ in
Eq. (\ref{eq:TN_decay}) for reversible binding.

Lawley found that the mean of the FPT $\Ti_{K,N}$ was determined for
any {\it fixed} $K$ as \cite{Lawley20b}
\begin{equation}
\langle \Ti_{K,N} \rangle \approx \frac{C}{\ln N}  \qquad (N\to \infty),
\end{equation}
with some constant $C$ and higher-order corrections $1/(\ln N)^2$
depending on $K$.  However, this behavior cannot be applied to $K =
N$.  In Appendix \ref{sec:Tmean_irrev}, we show that
\begin{equation}
\langle \Ti_{N,N} \rangle \approx C' \ln N  \qquad (N\to \infty),
\end{equation}
with another constant $C'$ determined by the decay time of the
survival probability for a single particle.  Even though the mean
arrival time of the slowest particle differs by a factor $(\ln N)^2$
from that of the fastest particle, the need for $N$ particles to
trigger the reaction event does not considerably slow down the
irreversible reaction kinetics.  This observation is totally different
in the case of reversible binding, for which the mean reaction time
$\langle \T_N\rangle$ in Eq. (\ref{eq:Tmean_asympt}) exhibits an
exponential growth with $N$.

\subsection{Comparison with the LMA}

Now, we compare our exact results to the Lawley-Madrid approximation.
This approximation was designed under assumption that the rebinding
time distribution can be approximated by an exponential law: $S(t)
\approx \bar{S}(t) = e^{-\nu t}$, with an appropriate rate $\nu$.
There are two natural choices for this rate.  In order to get the
correct long-time behavior of the survival probability, one can set
$\nu = D\lambda_1$ to match the leading term of the exact spectral
expansion (\ref{eq:S_spectral}).  Alternatively, as the rebinding time
$\tau$ is approximated by an exponential law, one can set $\nu =
1/\langle\tau \rangle$.  When the target is small and weakly reactive,
Eq. (\ref{eq:lambda1_small}) indicates that $1/\langle \tau \rangle$
is close to $D\lambda_1$, and both choices yield the same result.  One
sees that the approximate equality $1/\langle \tau \rangle \approx
D\lambda_1$ ensures the self-consistence of the Lawley-Madrid
approximation and can thus serve as a practical indicator of its
validity.  As a consequence, the LMA is expected to capture the
long-time behavior of the probability density $H_N(t|\x_0)$ in the
limit of small targets.  In the remaining part of this section, we
assume that the validity conditions of the LMA are fulfilled and set
$\nu = 1/\langle\tau\rangle$.

First, we look at the mean reaction time.  Lawley and Madrid analyzed
the asymptotic behavior of their Eq. (\ref{eq:LMA_mean}) in two
limits: (i) when $K$ is fixed and $N\to \infty$, in which case
$\langle \bar{\T}_{K,N}\rangle \sim \langle \tau\rangle K/N$, i.e.,
essentially a linear growth with $K$; (ii) when $1/(1+\koff \langle
\tau\rangle) < K/N < 1$ is fixed, in which case $\langle
\bar{\T}_{K,N}\rangle$ exhibits a very rapid growth \cite{Lawley19}.
While the limiting case $K = N$ was not discussed, we deduced the
asymptotic behavior of Eq. (\ref{eq:LMA_mean}) by using similar tools:
\begin{equation}  \label{eq:Tmean_LMA}
\langle \bar{\T}_{N,N}\rangle \approx \frac{(1 + \koff  \langle \tau\rangle)^N}{\koff N}  \quad (N\gg 1).
\end{equation}
This expression coincides with Eq. (\ref{eq:Tmean_asympt}) that we
obtained from the exact solution (\ref{eq:TN_mean}).  This highlights
that the LMA captures qualitatively the long-time behavior.  In turn,
as discussed earlier and illustrated in Table~\ref{tab:mean}, both
Eq. (\ref{eq:LMA_mean}) and its asymptotic form (\ref{eq:Tmean_LMA})
overestimate the mean reaction time.

Let us now turn to the approximation (\ref{eq:HKN_LMA}) of the
probability density.  Denoting by $0 > \nu_1 \geq \ldots \geq
\nu_N$ the negative eigenvalues of the matrix $W^{(N)}$, one sees that
the long-time asymptotic behavior is determined by the largest
eigenvalue $\nu_1$,
\begin{equation}
\bar{H}_{N,N}(t) \propto e^{-t/\bar{T}_N} \qquad (t\to\infty),
\end{equation}
with $\bar{T}_N = -1/\nu_1$.  In turn, the short-time approximation
reads
\begin{equation}
\bar{H}_{N,N}(t) \approx \frac{N}{\langle\tau\rangle} \, (t/\langle \tau\rangle)^{N-1} + O(t^{N})  \quad (t\to 0),
\end{equation}
where the lower-order powers of $t$ vanish because of the
three-diagonal structure of the matrix $W^{(N)}$, whereas the
prefactor in front of the leading term $t^{N-1}$ is
$\frac{[(W^{(N)})^{N-1}]_{N,1}}{(N-1)!} = N/\langle
\tau\rangle^{N-1}$.  Expectedly, this asymptotic behavior is different
from relation (\ref{eq:HN_short}) derived from our exact solution.
Note that in the limit $\koff \to 0$, the assumed exponential law for
the rebinding time implies
\begin{equation}
\lim\limits_{\koff\to 0} \bar{H}_{N,N}(t) = \frac{N}{\langle \tau \rangle} e^{-t/\langle\tau\rangle} (1 - e^{-t/\langle \tau\rangle})^{N-1}  .
\end{equation}

Figure \ref{fig:HN} illustrates the behavior of the probability
density $H_N(t|\x_0)$ for two values $N = 2$ and $N = 3$, as well as
two starting positions $|\x_0| = 5$ and $|\x_0| = 2$.  First of all,
we note that the probability density is broad, spanning over 4 to 6
orders of magnitude in time.  At short times, the probability density
$H_N(t|\x_0)$ does not depend on the unbinding rate $\koff$, yielding
the universal behavior of the left tail of the distribution given by
Eq. (\ref{eq:HKN}).  Note that the short-time asymptotic relation
(\ref{eq:HN_short}) is not accurate on the considered range of times
but captures correctly the leading-order term.  This relation can be
improved by including next-order corrections.  At the timescale
$1/\koff$, the unbinding mechanism starts to play a role, yielding
deviations from the irreversible binding case.  These deviations are
actually visible already at $t \gtrsim 5$ for $\koff = 0.03$ and $t
\gtrsim 50$ for $\koff = 0.003$.  As unbinding events slow down the
reaction, the right tail of the distribution is shifted towards longer
times as $\koff$ increases.  In fact, the long-time decay
(\ref{eq:HN_long}) is determined by the exponential function with the
decay time $T_N$ increasing with $\koff$.  Note that Monte Carlo
simulations are in perfect agreement with the exact solution.  

The comparison with the LMA reveals its advantages and limitations.
The LMA captures correctly the behavior of the probability density for
moderate and long times, the agreement being better as $\koff$ is
smaller.  One sees that the LMA systematically overestimates the decay
time that controls the long-time behavior (see Table~\ref{tab:mean}).
Deviations become larger as $N$ and $\koff$ increase.  Expectedly, the
LMA totally fails at short times.  Deviations are stronger when the
particles start closer to the target.  In fact, when $|\x_0| = 2$,
there is a notable maximum around $t \sim 1$ that is not captured by
the LMA.  The most probable time determining the position of this
maximum is several orders of magnitude smaller than the mean reaction
time.  This maximum can be relevant for applications when the source
of particles is close to the target (see \cite{Reva21} for further
discussions).
Nevertheless, the explicit character of the Lawley-Madrid
approximation and a much simpler computation of the probability
density via Eq. (\ref{eq:HKN_LMA}) make it a valuable tool for a
first-step analysis of reversible reactions with multiple particles.
Further improvements of the LMA present an important perspective.

\begin{figure*}
\begin{center}
\includegraphics[width=80mm]{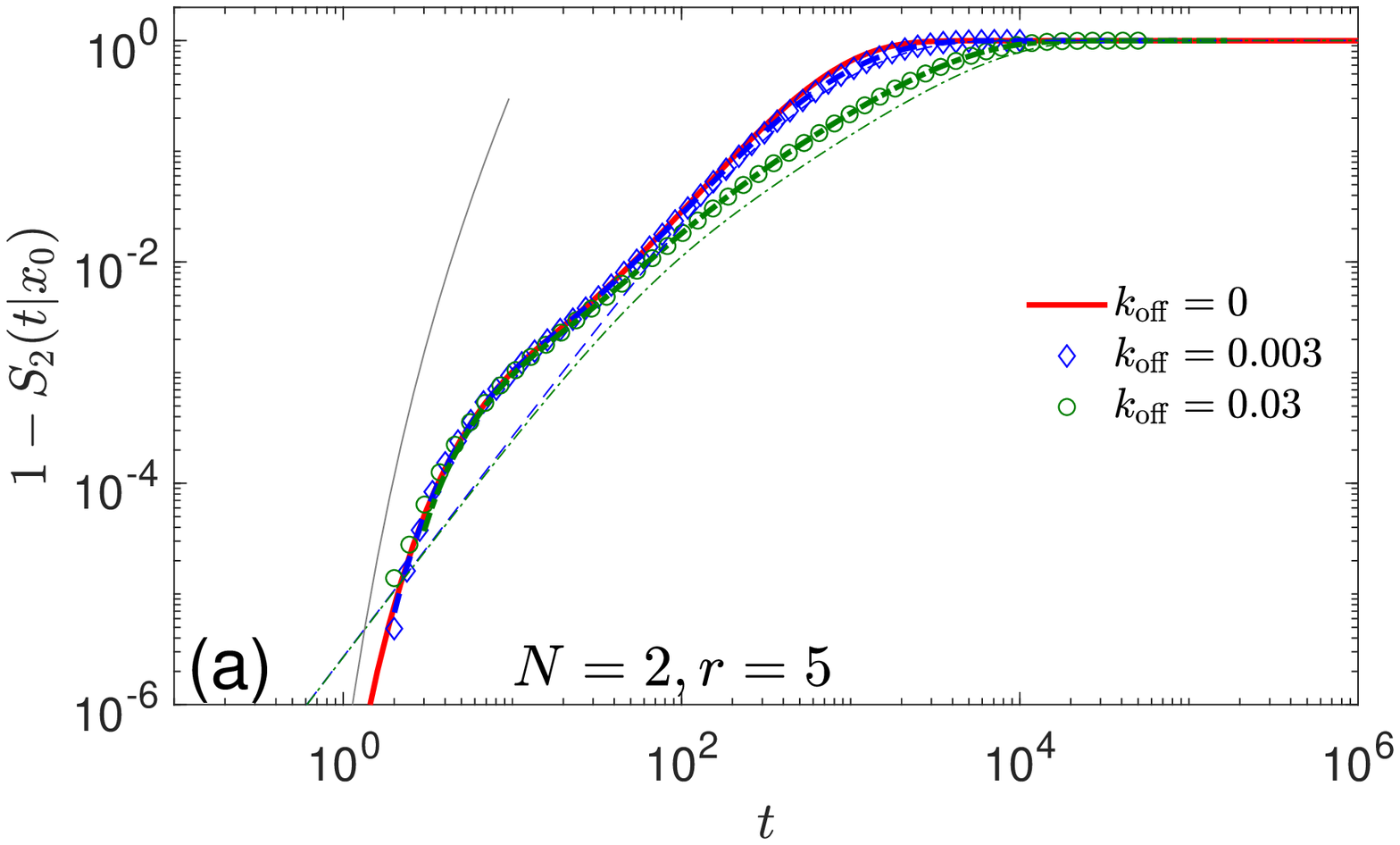} 
\includegraphics[width=80mm]{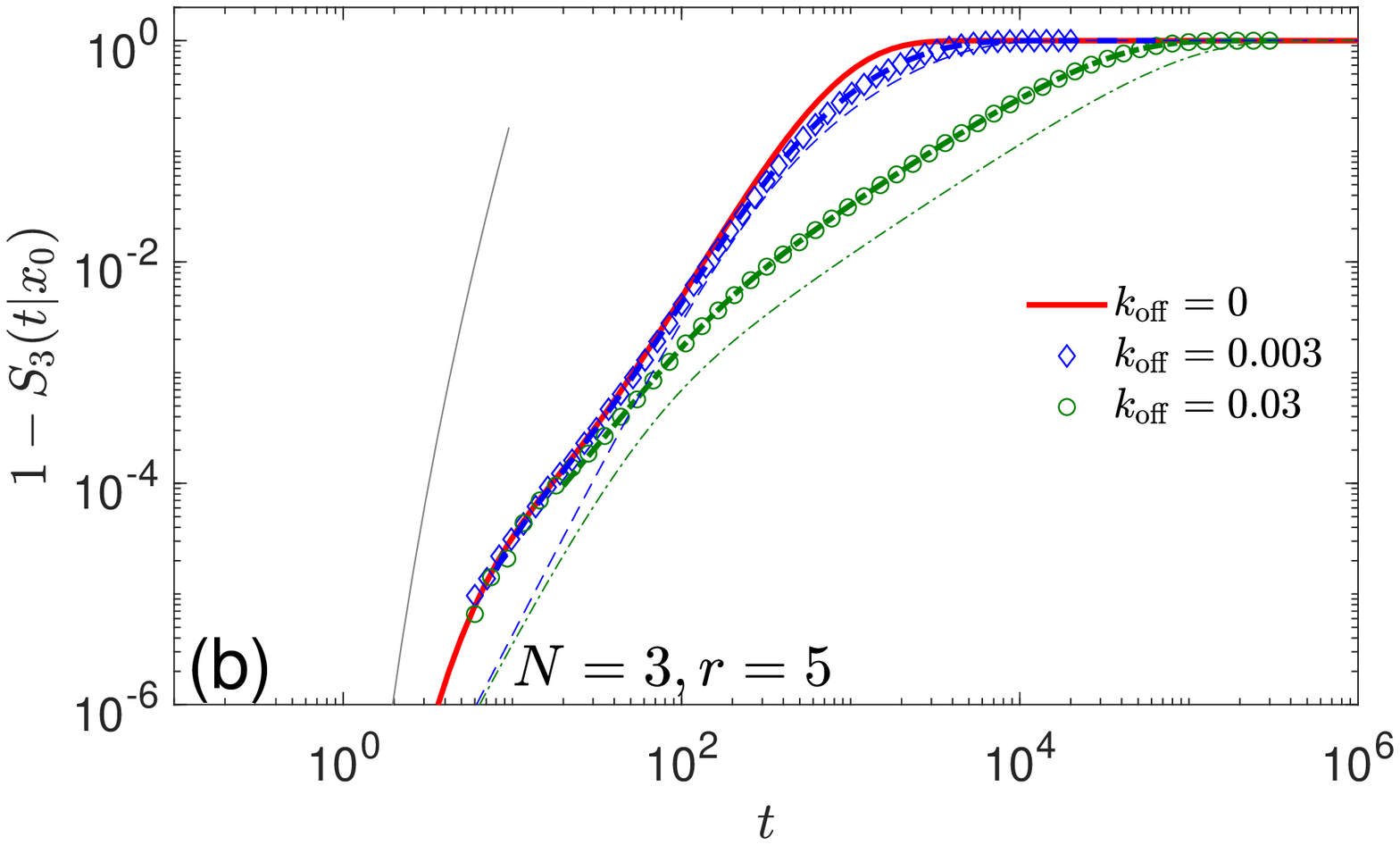} 
\includegraphics[width=80mm]{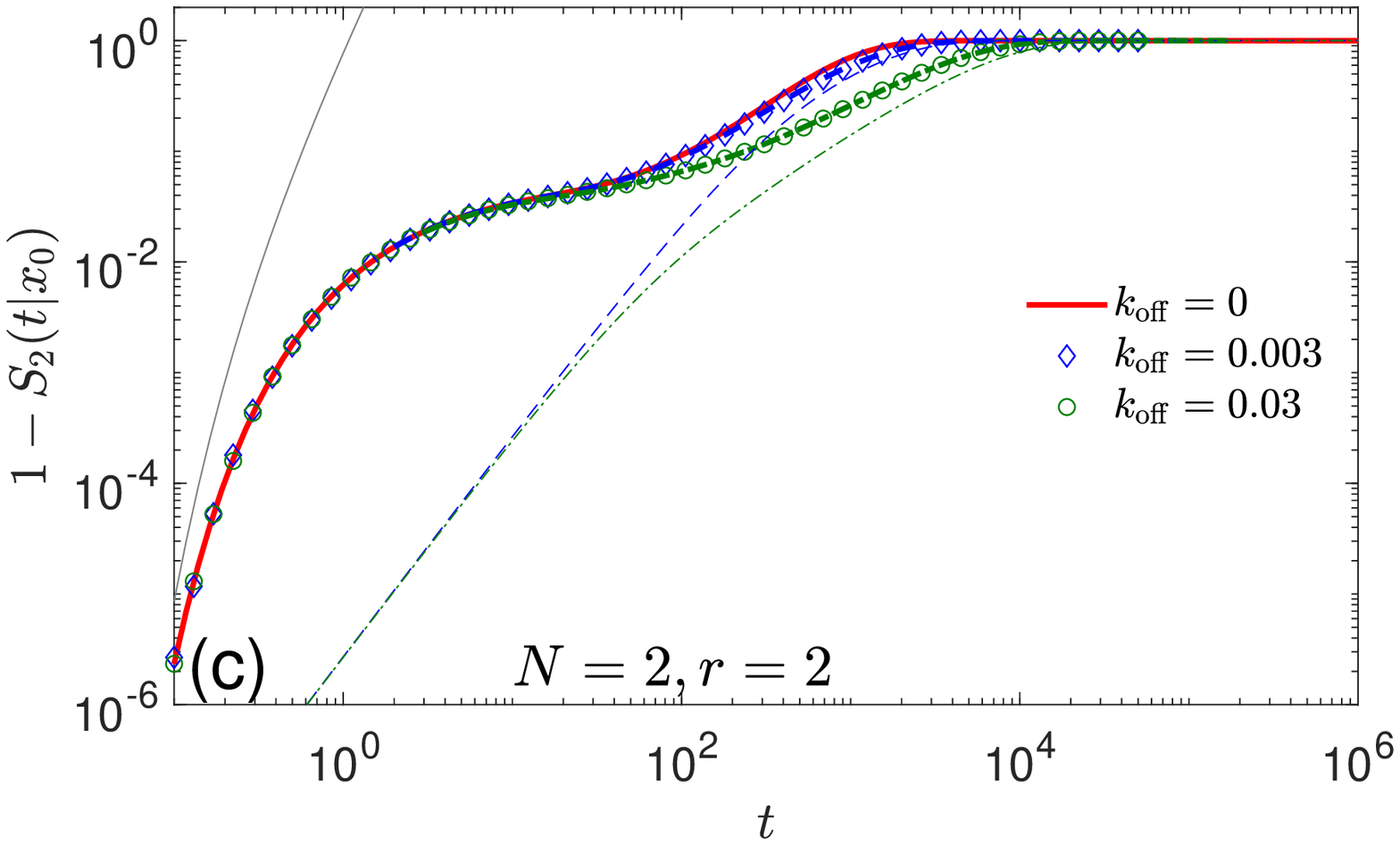} 
\includegraphics[width=80mm]{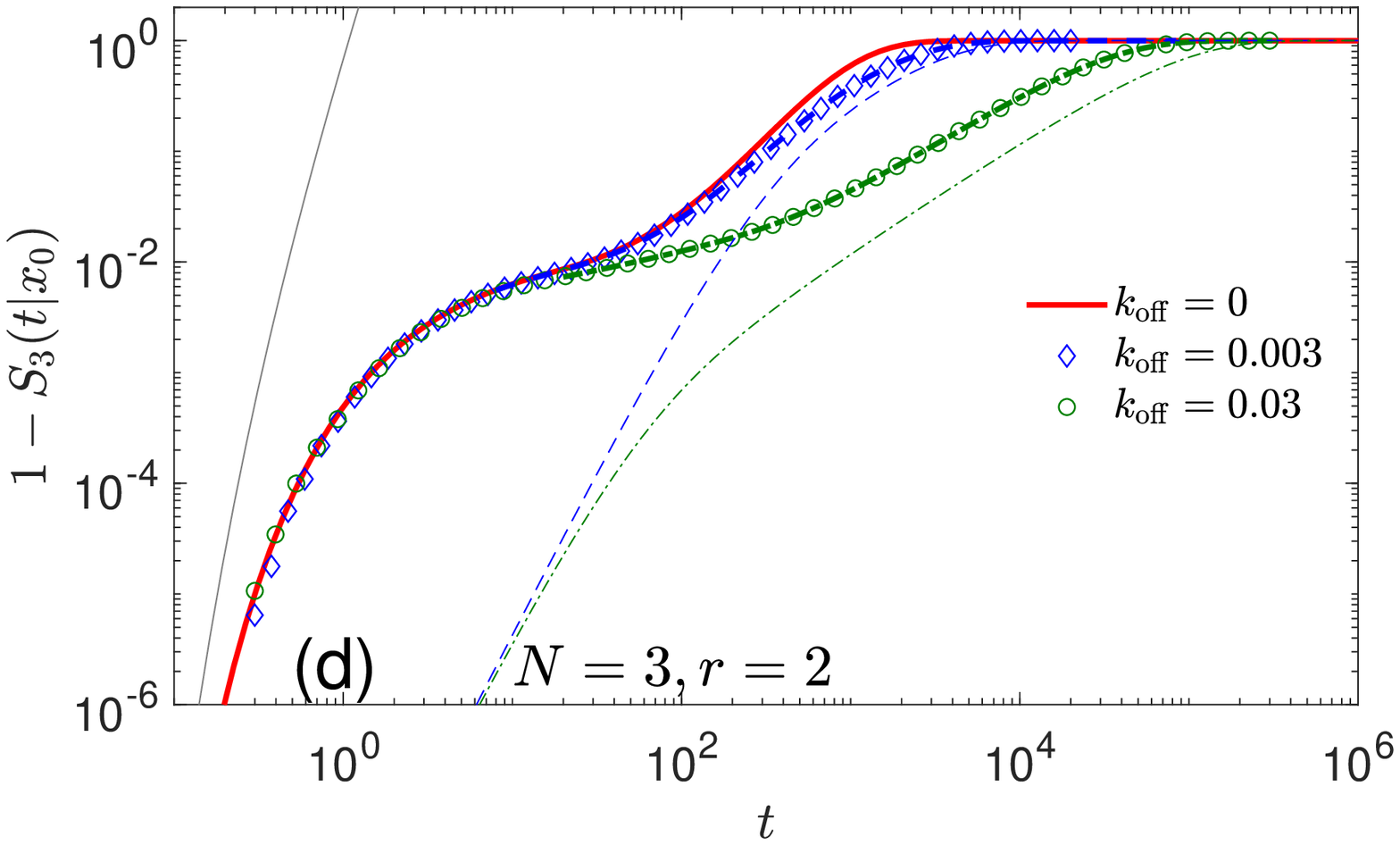} 
\end{center}
\caption{
Cumulative distribution function $\P\{\T_N < t\} = 1 - S_N(t|\x_0)$ of
the reaction time $\T_N$ for restricted diffusion between concentric
spheres of radii $\rho = 1$ and $R = 10$, with $D = 1$, $\kappa = 1$,
three values of $\koff$ (see legend), two starting positions $|\x_0| =
5$ {\bf (a,b)} and $|\x_0| = 2$ {\bf (c,d)}, and two values of $N$: $N
= 2$ {\bf (a,c)} and $N = 3$ {\bf (b,d)}.  Symbols show the empirical
cumulative distribution function of reaction times from Monte Carlo
simulations with $10^6$ particles.  Thick lines indicate the integral
of our exact solution (\ref{central}) evaluated numerically as
described in Appendix \ref{sec:computation}, whereas thin lines show
the Lawley-Madrid approximation (\ref{eq:SKN_LMA}).  Thin gray solid
line presents the short-time asymptotic behavior
(\ref{eq:SN_short}). }
\label{fig:SN}
\end{figure*}

Figure \ref{fig:SN} presents a complementary view onto the behavior of
the reaction time $\T_N$ by showing its cumulative distribution
function $1 - S_N(t|\x_0)$.

\section{Conclusions and perspectives}
\label{sec:conclusions}

Diffusion-controlled reactions involving multiple particles are
abundant and particularly relevant in biochemistry.  The need for a
sufficient number of bound particles can be considered as a sort of
protection mechanism against spontaneous triggering, as well as a mean
for reliable control of reactions.
The overwhelming majority of former studies in this field were focused
on first-passage times of a single particle, with a straightforward
extension to the extreme statistics of many particles with
irreversible binding to the target.  In turn, the problem of impatient
particles with reversible binding seems to remain unnoticed, in spite
of its practical relevance \cite{Grebenkov17}.  For instance, five
calcium ions have to bind to a calcium-sensing protein to initiate the
release of neurotransmitters for signalling between neurons
\cite{Berridge03,Eggermann12,Nakamura15,Dittrich13,Guerrier16,Reva21}.
To outline the role of unbinding kinetics onto this process, we take
the following estimates from Ref. \cite{Reva21}: $R = 300$~nm, $\kon =
6.35\cdot 10^8~\rm{M}^{-1}~\rm{s}^{-1} = 6.35\cdot
10^5~\rm{mol}~\rm{m}^{-3}~\rm{s}^{-1}$ and $\koff = 1.57\cdot
10^4~\rm{s}^{-1}$, from which the mean rebinding time is $\langle
\tau\rangle \approx 0.1$~s, see Eqs. (\ref{eq:mean_rebinding},
\ref{eq:eta}).  As a consequence, $\eta = \koff \langle\tau\rangle
\approx 1.57\cdot 10^3 \gg 1$, so that one cannot simply ignore
reversible binding that drastically changes the distribution of the
reaction time.

Even if the particles diffuse independently, their randomly
``asynchronized'' waiting times on the target render the problem of
exact characterization of the reaction times $\T_{K,N}$ mathematically
challenging.  The remarkable work by Lawley and Madrid brought an
elegant approximate solution to this problem \cite{Lawley19}.  The
good accuracy of this approximation, as reported by its authors, might
seem to suggest that this challenging problem is fully solved.  In
this paper, we showed that this is far from being the end of the
story.

We focused on the particular case of the first time $\T_N =
\T_{N,N}$ when all $N$ particles are bound to the target.  This choice
allowed us to derive, for the first time, the exact complete solution
of the problem of impatient particles, i.e., to express the
probability density of the random variable $\T_N$ in terms of the
first-passage time distribution of a single particle.  This exact
solution revealed some limitations and deficiencies of the LMA.  In
particular, we showed that the approximate solution captures the
qualitative behavior at moderate and long times but fails at short
times.  Moreover, the LMA overestimates the mean reaction time and the
decay time so that its predictions are inaccurate in some settings.
At the same time, the complexity of the exact solution for $\T_{N,N}$
and yet a fully open problem of finding the exact solution in the
general case $\T_{K,N}$ make the LMA a valuable tool for the
qualitative analysis and preliminary estimations.  Moreover, the
accuracy of the LMA is expected to be much higher in the limit of very
small targets.  We believe that further improvements of the LMA or
development of alternative methods can bring important insights on the
problem of impatient particles.  This is an interesting perspective of
the present work.

We also emphasize that impatient particles offer an excellent example
of a physical problem, for which standard numerical methods may be
insufficient for getting the whole picture.  In particular, as the
mean reaction time and the decay time grow exponentially fast with the
number of particles, getting the whole distribution of the reaction
time $\T_N$ was not possible even for moderate $N$.  For instance, a
Monte Carlo simulation with $10^6$ realizations used to plot the
empirical probability density in Fig. \ref{fig:HN} took one day on a
laptop.  However, this simulation allowed to get the behavior of
$H_N(t|\x_0)$ only for a limited range of time scales (e.g., from
$10^2$ to $10^4$ for $N = 2$).  Even though parallelization can easily
increase the number of realizations (say, by a factor 100 or 1000), it
would not be enough to get the short-time behavior.  As the
computational time explodes with $N$, we could not complete Monte
Carlo simulations even for moderate values of $N$ such as $N = 5$ or
$N = 10$.  Here, analytical tools and approximations are
indispensable.

While we focused on the setting when all particles start from the same
fixed point $\x_0$, our exact solution can be easily extended to a
more general case with distinct starting points.  Moreover, the
starting point of each particle can also be random.  In the case of a
uniform distribution of the starting points, the properties of the
fastest FPT $\Ti_{1,N}$ were studied in \cite{Grebenkov20}.  An
extension to the reaction time $\T_N$ is straightforward.  

The exact expression (\ref{eq:TN_mean}) for the mean reaction time
$\langle \T_N\rangle$ opens a way to investigate the role of different
parameters onto the reaction kinetics.  A rough approximation allowed
us to access the large-$N$ asymptotic behavior of this quantity.
However, the asymptotic formula (\ref{eq:Tmean_asympt}) lacks an exact
prefactor and also fails at small $N$.  More accurate analysis of
Eq. (\ref{eq:TN_mean}) could hopefully improve this formula to get a
quantitatively accurate description of the mean reaction time.  Its
extension to other reaction times $\langle \T_{K,N}\rangle$ presents
an exciting perspective.

\begin{acknowledgments}
DG acknowledges the Alexander von Humboldt Foundation for support
within a Bessel Prize award.  AK was supported by the Prime Minister's
Research Fellowship (PMRF) of the Government of India.
\end{acknowledgments}

\section*{Data Availability Statement}

The data that support the findings of this study are available from
the corresponding author upon reasonable request.

\appendix

\section{Mathematical details}
\label{sec:asymptotics}

In this Appendix, we discuss some asymptotic relations and derivations.

\subsection{Mean reaction time for irreversible binding}
\label{sec:Tmean_irrev}

The mean fastest FPT $\Ti_{1,N}$ and, more generally, the mean $K$-th
fastest FPT $\Ti_{K,N}$, were thoroughly investigated in the
irreversible binding case
\cite{Weiss83,Basnayake18,Basnayake19,Lawley20a,Lawley20b}.  For any
fixed $K$, the mean value $\langle \Ti_{K,N} \rangle$ behaves
universally as $1/\ln N$ in the large $N$ limit, whereas the
higher-order corrections $O(1/(\ln N)^2)$ depend on $K$.  In turn, the
asymptotic behavior of the slowest FPT $\Ti_{N,N}$ was not discussed,
to our knowledge.  In particular, the former result for any fixed $K$
cannot be applied to the case $K = N$.  Here, we sketch the main steps
of this analysis, more rigorous derivations being beyond the scope of
this paper.

\begin{figure}
\begin{center}
\includegraphics[width=85mm]{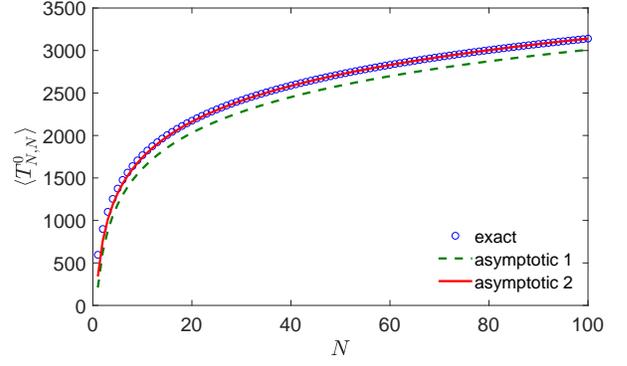} 
\end{center}
\caption{
Mean slowest FPT $\langle \Ti_{N,N}\rangle$ for restricted diffusion
between concentric spheres of radii $\rho = 1$ and $R = 10$, with
$|\x_0| = 5$, $D = 1$, $\kappa = 1$, and $\koff = 0$ (irreversible
binding).  Empty circles show the results of a numerical integration
of Eq. (\ref{eq:Tmean_irrev}), dashed line presents
Eq. (\ref{eq:TN_irrev_asympt}) with $\zeta = 0.5$, while solid line
illustrates Eq. (\ref{eq:TN_irrev_asympt}) with $\zeta = 0.428$, which
was selected to get the best agreement.}
\label{fig:Tmean_slow}
\end{figure}

As the mean time $\langle \Ti_{N,N}\rangle$ is given by
Eq. (\ref{eq:Tmean_irrev}), its asymptotic analysis is reduced to that
of the survival probability $S(t|\x_0)$ for a single particle.  It is
easy to check that the function $f(t) = 1 - (1-S(t|\x_0))^N$
monotonously decreases from $1$ at $t= 0$ to $0$ as $t\to\infty$.  The
integral in Eq. (\ref{eq:Tmean_irrev}) can be evaluated by
approximating $f(t)$ by the Heaviside step function $\Theta(t_N - t)$,
where $t_N$ is chosen by setting $f(t_N) = \zeta$, with $\zeta$ being
around $1/2$ (see below).  This equation yields $S(t_N|\x_0) = 1 -
(1-\zeta)^{1/N}$.  When $N$ is large, the right-hand side of this
relation is close to $0$.  In other words, the limit $N\to\infty$
corresponds to large $t_N$, for which the spectral expansion
(\ref{eq:S_spectral}) can be truncated to a single term, $S(t_N|\x_0)
\approx c_1 u_1(\x_0) e^{-D\lambda_1 t_N}$.  As a consequence, one gets
\begin{eqnarray}  \nonumber
\langle \Ti_{N,N}\rangle &\approx& t_N \approx - \frac{\ln (1-(1-\zeta)^{1/N}) - \ln(c_1 u_1(\x_0))}{D\lambda_1} \\  \label{eq:TN_irrev_asympt}
&\approx& \frac{\ln N + \ln(c_1 u_1(\x_0)) - \ln \ln \frac{1}{1-\xi}}{D\lambda_1} \,.
\end{eqnarray}
Even though this approximate relation depends on a somewhat arbitrary
choice of $\zeta$ around $1/2$, this dependence is weak and
corresponds to the sub-leading (constant) term, as compared to the
leading term $\ln N$.  Note that $1/(D\lambda_1)$ is the decay time
for a single particle which determines the natural timescale of the
problem.  Figure \ref{fig:Tmean_slow} illustrates the dependence of
$\langle \Ti_{N,N}\rangle$ on $N$ and its large-$N$ asymptotic
behavior (\ref{eq:TN_irrev_asympt}).

\subsection{Mean reaction time}
\label{sec:ATmean}

The mean reaction time $\langle \T_N\rangle$ is determined by
Eq. (\ref{eq:TN_mean}).  We note that the function $[Q(t)]^N -
[P(t|\x_0)]^N$ monotonously decreases from $1$ at $t = 0$ to $0$ as
$t\to\infty$.  We also checked that, for large $N$, this function
decreases fast enough to allow for truncation of the integral at some
finite time $t_N$, whereas the term $[P(t|\x_0)]^N$ is small for $t <
t_N$ and can be omitted.

\begin{figure}
\begin{center}
\includegraphics[width=85mm]{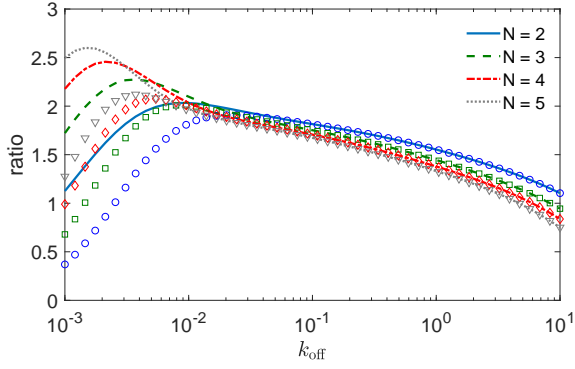} 
\end{center}
\caption{
The rescaled mean reaction time $\langle \T_N\rangle
\, \koff N P_{\infty}^{N}$ (lines) and the rescaled decay time $T_N
\, \koff N P_{\infty}^{N}$ (symbols) as functions of $\koff$ for
restricted diffusion between concentric spheres of radii $\rho = 1$
and $R = 10$, with $|\x_0| = 5$, $D = 1$, $\kappa = 1$, and $N = 2$
(circles and solid line), $N = 3$ (squares and dashed line), $N = 4$
(diamonds and dash-dotted line) and $N = 5$ (triangles and dotted
line).  Theoretical values of $\langle \T_N\rangle$ and $T_N$ were
obtained by numerical integration of Eqs. (\ref{eq:TN_mean}) and
(\ref{eq:TN_decay}), respectively. }
\label{fig:Tmean_asympt}
\end{figure}

One can thus apply the same approximation as in the previous
subsection.  In fact, we aim at evaluating $t_N$ at which $[Q(t_N)]^N
= \zeta$ or, equivalently, $Q(t_N) = \zeta^{1/N}$, with some $\zeta$
around $1/2$.  As $\zeta^{1/N}$ is close to $1$, one considers the
short-time approximation, for which $Q(t) = 1 - \koff t + O(t^2)$ (see
Eq. (\ref{eq:Qt}) and Appendix \ref{sec:Qt}).  We get thus $t_N =
(1-\zeta^{1/N})/\koff \approx \ln(1/\zeta)/(N\koff)$ as $N\to\infty$,
from which
\begin{equation}  \label{eq:TN_asympt0}
\langle \T_N \rangle \approx \frac{(1 + \koff \langle\tau\rangle)^N \ln (1/\zeta)}{\koff N} \qquad (N\to\infty),
\end{equation}
where we used Eq. (\ref{eq:Pinf}).  We stress that the prefactor
$\ln(1/\zeta)$ stands here in front of the leading term, whereas in
Sec. \ref{sec:Tmean_irrev}, an arbitrary parameter $\zeta$ appeared
only in the sub-leading term in Eq. (\ref{eq:TN_irrev_asympt}), while
the leading term was universal.  This feature highlights the
deficiency of the approximation (\ref{eq:TN_asympt0}).  Figure
\ref{fig:Tmean_asympt} compares the exact mean reaction time $\langle
\T_N\rangle$ and the asymptotic behavior $P_{\infty}^{-N}/(\koff N)$
for several values of $N$.  While the overall behavior is correctly
captured, deviations are considerable and depend on the parameters.
The curves shown in Fig. \ref{fig:Tmean_asympt} can be interpreted as
the dependence of $\ln(1/\zeta)$ on $\koff$ and $N$.  Further
improvements of this approximation present an interesting perspective.
Note also that the asymptotic behavior (\ref{eq:TN_asympt0}) with
$\ln(1/\zeta) = 1$ is identical to Eq. (\ref{eq:Tmean_LMA}) from the
LMA.

\begin{figure}
\begin{center}
\includegraphics[width=85mm]{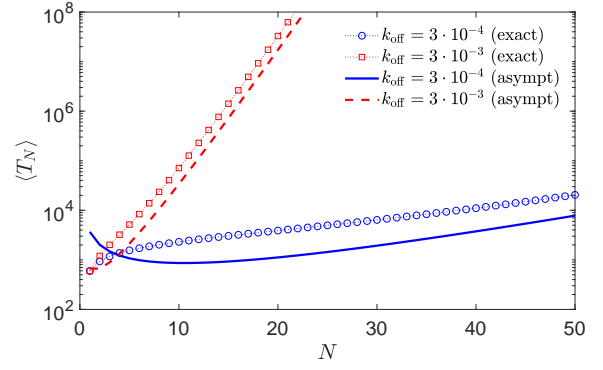} 
\end{center}
\caption{
Mean reaction time $\langle \T_N\rangle$ (symbols) and its asymptotic
form (\ref{eq:Tmean_asympt}) as functions of $N$ for restricted
diffusion between concentric spheres of radii $\rho = 1$ and $R = 10$,
with $|\x_0| = 5$, $D = 1$, $\kappa = 1$, and two values of $\koff$.
Theoretical values of $\langle \T_N\rangle$ were obtained by numerical
integration of Eq. (\ref{eq:TN_mean}). }
\label{fig:Tmean_asympt_opt}
\end{figure}

In a first approximation, one may attempt to set the factor
$\ln(1/\zeta)$ to $1$, as in Eq. (\ref{eq:Tmean_LMA}).  The
non-monotonous dependence of the right-hand side of the asymptotic
form (\ref{eq:TN_asympt0}) on $\koff$ and $N$ may suggest that the
mean reaction time can be optimized with respect to these parameters.
In fact, its derivative with respect to $N$ vanishes at
\begin{equation}
N_c = \frac{1}{\ln(1 + \koff \langle\tau\rangle)} \,,
\end{equation} 
suggesting that $\langle \T_N\rangle$ can be minimized with respect to
$N$ when $\koff \langle\tau\rangle$ is small enough.  Similarly, the
derivative with respect to $\koff$ vanishes at
\begin{equation}
k_{\rm off,c} = \frac{1}{(N-1)\langle\tau\rangle} \,,
\end{equation}
suggesting a minimum of $\langle \T_N\rangle$.  However, this
fictitious optimality results from a rough asymptotic formula
(\ref{eq:TN_asympt0}) and does not occur when the exact solution
(\ref{eq:TN_mean}) is considered (see
Fig. \ref{fig:Tmean_asympt_opt}).  This example illustrates danger of
relying on approximate solutions and urges for a more elaborate
analysis of the exact solution.

\subsection{The decay time for $N=1$}
\label{sec:caseN1}

As mentioned in Sec. \ref{sec:long}, the approximation
(\ref{eq:TN_decay}) of the decay time $T_N$ is least accurate in the
case $N = 1$.  To illustrate this point, we note that the integral in
Eq. (\ref{eq:TN_decay}) can be found explicitly for $N = 1$.  In fact,
one has
\begin{eqnarray*}
&& \int\limits_0^\infty dt \bigl(Q(t) - P_\infty\bigr) = \lim\limits_{p\to0} \biggl(\tilde{Q}(p) - \frac{P_\infty}{p}\biggr) \\
&& = \lim\limits_{p\to 0} \biggl(\frac{1}{p + \koff(1-\tilde{H}(p))} - \frac{P_{\infty}}{p}\biggr) \\
&& = \lim\limits_{p\to 0} \biggl(\frac{1}{p + \koff(p\langle\tau\rangle - p^2 \langle \tau^2/2 \rangle + O(p^3))} - \frac{P_{\infty}}{p}\biggr) \\
&& = \frac12 P_\infty^2 \koff \langle\tau^2\rangle  ,
\end{eqnarray*}
where we used Eq. (\ref{eq:Pinf}).  As a consequence, the
approximation (\ref{eq:TN_decay}) reads
\begin{equation}
p_{1,1} \approx - \frac{2}{P_{\infty} \koff \langle\tau^2\rangle} \,.
\end{equation}
Note that for restricted diffusion between concentric spheres, the
second moment of the rebinding time is known explicitly, see
Eq. (\ref{eq:tau2}).  However, a similar approximation can be used to
estimate the first pole of $\tilde{Q}(p)$, by expanding $\tilde{H}(p)$
in Eq. (\ref{eq:Qtilde}) up to the second order in $p$, from which
$p_1 \approx - 2/(P_\infty \koff \langle \tau^2\rangle)$.  In other
words, $p_{1,1}$ turns out to be identical to $p_1$, thus invalidating
the approximation in the case $N = 1$.

\subsection{Behavior of the function $Q(t)$}
\label{sec:Qt}

According to Eq. (\ref{eq:Qt}), $Q(t)$ is a monotonously decreasing
function.  In fact, the time derivative of Eq. (\ref{eq:Qt}) reads
\begin{equation}
Q'(t) = - \koff \bigl(Q(t) - P(t)\bigr).
\end{equation}
Comparing the probabilities $P(t)$ and $Q(t)$ of finding the particle
bound to the target, one realizes that the former includes an
additional step of binding to the target and thus $P(t) \leq Q(t)$,
implying $Q'(t) \leq 0$.  This property can also be deduced in a more
formal way.  In fact, as $P(t)$ is the convolution of $Q(t)$ and
$H(t)$ (see Eq. (\ref{eq:Ptilde_Qtilde})), its integration by parts
yields
\begin{eqnarray*}
P(t) &=& \int\limits_0^t dt' \, Q(t-t') H(t') \\
&=& - \bigl[Q(0) S(t) - S(0) Q(t)\bigr] + \int\limits_0^t dt' \, Q'(t-t') S(t') .
\end{eqnarray*}
Since $S(0) = Q(0) = 1$, one deduces 
\begin{equation}
Q'(t) = - \koff \biggl(S(t) - \int\limits_0^t dt' \, S(t') Q'(t-t')\biggr),
\end{equation}
where $S(t) \geq 0$ is the survival probability.  Applying a sort of
induction argument, one can check that the right-hand side is
negative.  Note also that this relation implies $Q'(0) = -\koff$ and
thus $Q(t) \approx 1 - \koff t + O(t^2)$ as $t\to 0$.

\section{Diffusion between concentric spheres}
\label{sec:Aspheres}

In this Appendix, we summarize former results needed for evaluating
the probability density of the reaction time $\T_N$ for the
practically relevant scenario of particles diffusing in a shell-like
domain $\Omega=\{ \x \in \R^3: \rho< | \x |<R\}$ bounded between two
concentric spheres of radii $\rho$ and $R$.  The inner sphere is a
partially reactive target with reactivity $\kappa$, whereas the outer
sphere is reflecting.  The rotational symmetry of the problem allows
for an explicit solution by separation of variables
\cite{Redner,Gardiner,Carslaw}.  The first-passage time distribution
was discussed in \cite{Grebenkov18}, whereas the exact solution for
the probability $P(t|\x_0)$ was given in \cite{Reva21}.

\subsection{First-passage time density}

The probability density of the first-passage time can be found by
separation of variables in a standard way (see \cite{Grebenkov18} for
details).  The rotational symmetry implies that $H(t|\x_0)$ and other
related quantities depend only on time $t$ and the radial coordinate
$r = |\x_0|$.  In Laplace domain, one has
\begin{equation}  \label{eq:Hp}
\tilde{H}(p|\x_0) = \frac{g(r)}{g(\rho)-g'(\rho)\frac{D}{\kappa}} \,,
\end{equation}
where 
\begin{equation}  \label{eq:gr}
g(r)=\frac{R\sqrt{p/D}\cosh\xi-\sinh\xi}{r\sqrt{p/D}} \,,
\end{equation}
with $\xi=(R-r)\sqrt{p/D}$, and $g'(r)$ is given by
\begin{equation}  \label{eq:dgr}
g'(r)=\frac{(1-Rr p/D)\sinh\xi-\xi\cosh\xi}{r^2\sqrt{p/D}} \,.
\end{equation}
The moments of the first-passage time can be found as
\begin{equation}
\langle \tau^k\rangle_{\x_0} = (-1)^k \lim\limits_{p\to 0} \frac{\partial^k \tilde{H}(p|\x_0)}{\partial p^k} \,.
\end{equation}
Setting $|\x_0| = \rho$, one also determines the moments of the
rebinding time, e.g.,
\begin{equation} 
\langle \tau\rangle = \frac{R^3-\rho^3}{3\kappa \rho^2} 
\end{equation}
and
\begin{eqnarray}  \nonumber
\langle \tau^2\rangle &=& \frac{2(R^3-\rho^3)^2}{9 \kappa^2 \rho^4} \\  \label{eq:tau2}
&+& \frac{2(5R^6 - 9R^5\rho + 5R^3\rho^3 - \rho^6)}{45 D\kappa \rho^3} \,.
\end{eqnarray}

The inversion of the Laplace transform in Eq. (\ref{eq:Hp}) by means
of the residue theorem yields:
\begin{equation}  \label{eq:Ht}
H(t|\x_0) = \frac{D}{\rho^2} \sum\limits_{n=1}^\infty \hat\alpha_n^2 \, \hat c_n \, u\bigl(\hat\alpha_n, |\x_0|\bigr) \, e^{- Dt \hat\alpha_n^2/\rho^2} ,
\end{equation}
with
\begin{eqnarray}  \label{eq:un}
&& u(\alpha, r) = \frac{\rho \, \sin\bigl(\alpha \frac{R-r}{\rho}\bigr) - R\alpha \cos\bigl(\alpha \frac{R-r}{\rho}\bigr)}{r} \,, \\
\nonumber
&& \hat c_n = - \frac{2\mu \rho^2}{\hat\alpha_n} \biggl[(\mu R(R-\rho) + R^2+\rho^2)\\ \nonumber 
&& \times \hat\alpha_n\sin(\hat\alpha_n\beta) + (R(R-\rho)\hat\alpha_n^2 - \mu \rho^2)\cos(\hat\alpha_n \beta)\biggr]^{-1} \,,
\end{eqnarray}
and $\hat\alpha_n$ (with $n=1,2,\ldots$) denoting the positive solutions
of the trigonometric equation
\begin{equation} \label{eq:eq_alpha}
\tan(\alpha \beta) = \frac{\alpha\bigl(\beta + (1+\beta)\mu\bigr)}{1+\mu + (1+\beta)\alpha^2}  \,,
\end{equation}
with 
\begin{equation}
\mu=\kappa \rho/D, \qquad \beta = (R-\rho)/\rho.
\end{equation}
Note that the survival probability is obtained by integrating
Eq. (\ref{eq:Ht}):
\begin{equation}  \label{eq:St_sphere}
S(t|\x_0) = \sum\limits_{n=1}^\infty \hat c_n \, u\bigl(\hat\alpha_n, |\x_0|\bigr) \, e^{- Dt \hat\alpha_n^2/\rho^2} .
\end{equation}

\subsection{The occupancy probability}

In turn, the spectral expansion (\ref{eq:P_spectral}) of the occupancy
probability $P(t|\x_0)$ was derived in \cite{Reva21}, with $v_n(\x_0)
= c_n u(\alpha_n,|\x_0|)$, where $u(\alpha,r)$ is given by
Eq. (\ref{eq:un}), $p_n = - \alpha_n^2 D/\rho^2$, and
\begin{eqnarray*}  
c_n &=& \frac{2 \mu}{\sin(\alpha_n \beta) \bigl(\alpha_n^2 w_1 + w_2 \bigr) 
 + \alpha_n \cos(\alpha_n\beta) \bigl(\alpha_n^2 w_3 + w_4 \bigr)} ,
\end{eqnarray*}
with
\begin{subequations}
\begin{align}
w_1 & = 4(1+\beta) + \beta(\beta+\mu(1+\beta)) , \\
w_2 & = 2(1+\mu-\lambda(1+\beta)) - \lambda \beta^2 , \\
w_3 & = \beta(1+\beta) ,\\
w_4 & = \beta(1 + \mu - \lambda(1+\beta)) - 3(\beta + \mu(1+\beta))  ,
\end{align}
\end{subequations}
$\lambda = \koff \rho^2/D$, and $\alpha_n$ are strictly positive
solutions of the trigonometric equation
\begin{equation}
\label{eq:alpha_M0}
\sin(\alpha_n \beta) = \frac{\bigl[\alpha_n^2(\beta+\mu(1+\beta)) - \lambda \beta\bigr] \alpha_n \cos(\alpha_n \beta)}
{\alpha_n^4(1+\beta) + \alpha_n^2(1+\mu - \lambda(1+\beta)) - \lambda}  \,.
\end{equation}
enumerated by $n=1,2,\ldots$.  Note that the coefficients $v_n$
determining $Q(t)$ in Eq. (\ref{eq:Qt_spectral}) are simply $v_n = c_n
\, u(\alpha_n,\rho)$.

\subsection{Short-time asymptotic behavior}
\label{sec:spheres_short}

Here we focus on the short-time behavior of $H(t|\x_0)$.  As the
solution in Eq. (\ref{eq:Ht}) depends only on the radial coordinate $r
= |\x_0|$, we replace $\x_0$ by $r$ in the following expressions.
Setting $s = (R-\rho)^2p/D$ and $\nu = D/(\kappa \rho)$, we can
rewrite $\tilde{H}(p|r)$ as
\begin{eqnarray}  \label{eq:Hs}
&& \tilde{H}(p|r) = \frac{\beta\sqrt{s}\cosh{\big(\sqrt{s}\frac{R-r}{R-\rho}\big)}-\sinh{\big(\sqrt{s}\frac{R-r}{R-\rho}\big)}}{r/\rho}\\  \nonumber
&& \times \bigg( (\nu + \beta)\sqrt{s}\cosh(\sqrt{s})-(1+\nu-\gamma s)\sinh(\sqrt{s})\bigg)^{-1},  
\end{eqnarray}
where $\gamma = \nu \beta^2\rho/R$.  We first study the case
$r=\rho$, for which
\begin{equation}
    \tilde{H}(p|\rho) = \frac{1-\frac{1}{\beta\sqrt{s}}\tanh(\sqrt{s})}{1+\nu/\beta -\frac{1+\nu-\gamma s}{\beta\sqrt{s}}\tanh(\sqrt{s})} \,.
\end{equation}
For large $s$, $\tanh(\sqrt{s}) = 1+O(e^{-2\sqrt{s}})$, which further
implies
\begin{eqnarray*}
    \tilde{H}(p|\rho) &\approx& \frac{\beta\sqrt{s}-1}{(\nu+\beta)\sqrt{s}-(1+\nu)+\gamma s} \\
&=& \frac{1}{\gamma}\biggr( \frac{A_1}{\sqrt{s}+x_1} + \frac{A_2}{\sqrt{s}+x_2}\bigg),
\end{eqnarray*}
where
\begin{equation}
    x_{1,2} = \frac{(\nu+\beta)\pm \sqrt{(\nu+\beta)^2+4\gamma(1+\nu)}}{2\gamma} \,,
\end{equation}
and 
\begin{equation}
A_{1} = \frac{1+\beta x_1}{x_{1}-x_{2}} \,, \qquad
A_{2} = \frac{1+\beta x_2}{x_{2}-x_{1}} \,.
\end{equation}
Using the following inverse Laplace transform, 
\begin{equation}
    \mathcal{L}\biggr\{ \frac{1}{\sqrt{s}+ a} \biggr\}= \frac{1}{\sqrt{\pi \tau}} - ae^{a^2\tau}\text{erfc}(a\sqrt{\tau})=f_a(\tau)
\end{equation}
(where $\textrm{erfc}(z)$ is the complementary error function), we
find the short-time approximation
\begin{equation} 
    H(t|\rho) \approx D \frac{A_1 f_{x_1}\big(\frac{Dt}{(R-\rho)^2}\big) + A_2 f_{x_2}\big(\frac{Dt}{(R-\rho)^2}\big)}{(R-\rho)^2\gamma} \,.
\end{equation}
Using $f_a(\tau)\approx 1/\sqrt{\pi \tau} -a +
2a^2\sqrt{\tau/\pi}+O(\tau)$, one gets
\begin{eqnarray} \nonumber
    H(t|\rho) &\approx & \frac{\kappa}{\sqrt{\pi D t}} + \kappa(1/\rho + \kappa/D) \\   \label{hrhot_f}
&+& 2\frac{\kappa(1/\rho + \kappa/D)^2}{\sqrt{\pi}}\sqrt{Dt}+O(t).
\end{eqnarray}    
For the case of $\rho < r < R$, we have
\begin{equation}
    \tilde{H}(p|r) \approx \frac{\rho}{r} e^{-\sqrt{s}(r-\rho)/(R-\rho)} \tilde{H}(p|\rho) ,
\end{equation}
where we neglected the terms of the order $e^{-2\sqrt{s}}$ and
$e^{-2\sqrt{s}\delta}$, with $\delta = (R-r)/(R-\rho)$. Note that if
$r$ is close to $R$ (i.e., if $\delta$ is very small), the above
approximation would be slightly modified.  Since the Laplace transform
is expressed as a product of two terms, the inverse Laplace transform
yields the convolution
\begin{equation}
    H(t|r) \approx \frac{\rho}{r}\int_0^t dt' H(t-t'|\rho) \frac{(r-\rho)e^{-(r-\rho)^2/(4Dt')}}{\sqrt{4\pi D t'^3}} \,,
\end{equation}
which can be evaluated using the asymptotic expression for
$H(t|\rho)$ to give
\begin{equation}  \label{hrt_short_spheres}
 H(t|r) = \frac{\rho\kappa e^{-(r-\rho)^2/(4Dt)}}{r\sqrt{\pi Dt} } \bigg( 1+ \frac{2Dt(1+\kappa \rho/D)}{\rho(r-\rho)}+ \cdots \bigg). 
\end{equation}
One can recognize Eq. (\ref{hrt_short}) in the leading term, with
$\alpha = -1/2$ and
\begin{equation}
C_{\x_0} = \frac{\rho\kappa}{r\sqrt{\pi D}} \,.
\end{equation}

\section{Numerical implementation}
\label{sec:computation}

Our central formula \eqref{central} expresses the probability density
$H_N(t|\x_0)$ in terms of the accessible probabilities $P(t|\x_0)$ and
$Q(t)$.  However, its practical implementation requires the
computation of two Laplace transforms, $\L\{[P(t|\x_0)]^N\}$ and
$\L\{[Q(t)]^N\}$, and then the evaluation of the inverse Laplace
transform of their ratio.  Since both $P(t|\x_0)$ and $Q(t)$ are given
as spectral expansions, such a computation becomes numerically
difficult, especially at small and large times when $H_N(t|\x_0)$
rapidly decays.  We also attempted a direct solution of the related
deconvolution problem:
\begin{equation}  \label{eq:deconv0}
[P(t|\x_0)]^N = \int\limits_0^t dt' \, H_N(t'|\x_0) \, [Q(t-t')]^N ,
\end{equation}
but it was numerically unstable.  

To resolve this difficulty, one can integrate Eq. (\ref{eq:deconv0})
by parts to transform it into an integral equation on the survival
probability $S_N(t|\x_0) = \P\{\T_N > t\}$:
\begin{eqnarray}  \label{eq:auxil11}
[Q(t)]^N - [P(t|\x_0)]^N &=& S_N(t|\x_0) \\     \nonumber
&-& \int\limits_0^t dt' \, S_N(t'|\x_0) \, q_N(t-t') ,
\end{eqnarray}
where we used $Q(0) = S_N(0|\x_0) = 1$, and defined
\begin{equation}
q_N(t) = - \frac{d}{dt} [Q(t)]^N = -N [Q(t)]^{N-1} \frac{dQ}{dt} \,.
\end{equation}
Considering the last term in Eq. (\ref{eq:auxil11}) as the application
of an integral operator $\Q$ to the function $S_N(t|\x_0)$, one can
formally invert this relation to get
\begin{equation}
S_N(t|\x_0) = (I - \Q)^{-1} \bigl([Q(t)]^N - [P(t|\x_0)]^N\bigr),
\end{equation}
where $I$ is the identity operator.  Expanding the operator
$(I-\Q)^{-1}$ into the geometric series, one finally expresses the
survival probability as
\begin{equation}    \label{eq:SN_conv}
S_N(t|\x_0) = \int\limits_0^t dt' \bigl([Q(t')]^N - [P(t'|\x_0)]^N\bigr) R(t-t') ,
\end{equation}
where
\begin{eqnarray}  
R(t) &=& (I - \Q)^{-1} \delta(t) \\  \nonumber
&=& \delta(t) + q_N(t) + \int\limits_0^t dt_1 \, q_N(t_1) \, q_N(t-t_1) + \ldots ,  
\end{eqnarray}
i.e., the sum of convolutions of $q_N(t)$ with itself of all orders.

In practice, we compute both $[Q(t)]^N - [P(t|\x_0)]^N$ and $q_N(t)$
over a linear grid of points $\{0,\delta,2\delta,\ldots, \delta
(K-1)\}$ and then evaluate convolutions by fast Fourier transform
(FFT).  In this way, one gets the survival probability evaluated at
grid points:
\begin{equation}
S_N(j\delta|\x_0) \approx \F^{-1} \left\{ \frac{\F\{ p_k \}}{1 - \F\{ q_k \}} \right\} \,,
\end{equation}
with $j = 0,1,\ldots,K-1$, where $p_k = c_k \bigl([Q(k\delta)]^N -
[P(k\delta|\x_0)]^N\bigr)$, $q_k = \delta c_k q_N(k\delta)$, $c_0 =
1/2$, $c_k = 1$ for $0 < k < K$, and $c_k = 0$ for $K \leq k < 2K$.
Here the coefficient $c_0$ accounts for the integration weight $1/2$
of the first point, whereas $c_k$ for $k \geq K$ allow one to pad the
vectors by $0$ for the proper computation of linear convolutions via
direct ($\F$) and inverse ($\F^{-1}$) FFTs applied to vectors of
length $2K$.  Note that the probability density $H_N(t|\x_0)$ can also
be found via FFT as
\begin{equation}
H_N(j\delta |\x_0) \approx \frac{1}{\delta} \F^{-1} \left\{ \bigl(e^{2\pi ik/(2K)}-1\bigr) \frac{\F\{ p_k \}}{1 - \F\{ q_k \}} \right\} \,,
\end{equation}
in analogy with the evaluation of a derivative via standard Fourier
transform: $f'(x) = \F^{-1} \{ ik \F\{ f(x)\}\}$.

The time step $\delta$ sets the minimal time at which both
$S_N(t|\x_0)$ and $H_N(t|\x_0)$ are available, and controls the
accuracy of the whole computation.  In fact, it determines how
accurately discrete sums approximate convolution integrals.  This is
particularly important for the evaluation of $\F\{q_k\}$, whose
maximal value is achieved at the zero frequency:
\begin{eqnarray}  \nonumber
\max\{ |\F\{q_k\}|\} &=& \F_0\{q_k\} = \sum\limits_{k=0}^{2K-1} q_k \approx  \int\limits_0^{t_{\rm max}} dt \, q_N(t) \\  \nonumber
&=& 1 - [Q(t_{\rm max})]^N \approx 1 - P_\infty^N < 1.
\end{eqnarray}
As a consequence, $1/(1 - \F\{ q_k \})$ is well defined.  However,
when $N$ or $\koff$ increase, the maximum approaches to $1$.  If
$\delta$ is not small enough, inaccurate discretization may result in
$\F_0\{q_k\}$ exceeding $1$ and thus strong instabilities in the above
computation.  For the computation of theoretical curves in
Fig. \ref{fig:HN}, we used $\delta = 0.01$ in all cases, except for
the case $\koff = 0.03$ and $N=3$, for which $\delta = 0.005$ was
needed.

\section{Monte Carlo simulations}
\label{sec:MC}

Monte Carlo simulations were realized via a standard event-driven
scheme.  Each particle was equipped by its internal ``clock'' $t_i$
and the binary state variable $s_i$ indicating whether the particle is
bound or not.  At time $0$, all particles are free ($s_i = 1$) and
released from a fixed point $\x_0$, with their clocks being set to
$0$.  The particles diffuse independently and bind the target at
random times sampled from the probability density $H(t|\x_0)$.  The
internal clock of each particle is thus set to its (individual)
first-binding time, while their states are set to $0$ (bound).  We
emphasize that these FPTs account for partial reactivity of the
target, i.e., for eventual failed binding attempts and reflections
from the target, until the successful binding.  Selecting the particle
with the minimal internal clock (say, $t_i$), one updates this clock
by adding a random waiting time $\delta_i$ generated from the
exponential law with the rate $\koff$, and sets its state variable
$s_i$ to $1$ (free).  In other words, $t_i$ is replaced by
$t_i+\delta_i$, which is the instance when the $i$-th particle unbinds
from the target and resumes its diffusion.  From now on, the following
step is repeated: one selects the particle with the minimal internal
time (say, $t_j$); if $s_j = 0$ (i.e., at the instance $t_j$ the
particle binds to the target), we evaluate the number of bound
particles at time $t_j$, and the simulation is stopped if all
particles are bound; if the simulation is not stopped, the clock $t_j$
is updated by adding either a newly generated random waiting time
$\delta_j$ (if $s_j = 0$), or a random rebinding time $\tau$ sampled
from the probability density $H(t)$ (if $s_j = 1$).  This step is
repeated until the simulation is stopped (see Fig. \ref{fig:Nt}).

The first-binding times are generated from the known probability
density $H(t|\x_0)$ given by Eq. (\ref{eq:Ht}).  To sample from a
broad distribution $H(t|\x_0)$ spanning several orders of magnitude in
time, we first perform a change of variable $\zeta = \ln t$, and
obtain the associated probability density $H_1(\zeta|\x_0)$.  Prior to
running simulations, we create a linear grid of possible values
$\zeta_k$, ranging from $\zeta_{\min}$ to $\zeta_{\rm max}$, with a
step $d\zeta = 0.01$, and a grid containing the probability weight
$H_1(\zeta_k|\x_0) d\zeta$ of each value $\zeta_k$.  Using these
probability weights, a (pseudo)-random value of $\zeta$ is generated
by using the Matlab function \verb|randsample|, and the corresponding
first-binding time is obtained as $e^\zeta$.  The same method is used
for generating rebinding times from the known probability density
$H(t)$.

For the considered example of restricted diffusion between two
spheres, the explicit form of the survival probability $H(t|\x_0)$ is
provided in Appendix \ref{sec:Aspheres}.  The spectral decomposition
(\ref{eq:St_sphere}) was truncated at a large order $n = 10000$ in
order to access accurately the short-time behavior of $H(t|\x_0)$.
The zeros $\hat{\alpha}_n$ of Eq. (\ref{eq:eq_alpha}) were found by
the bisection method (see \cite{Grebenkov18,Reva21} for details).  The
grid bounds $\zeta_{\rm min}$ and $\zeta_{\rm max}$ depend on the
parameters and were chosen manually to cover a broad range of times
whose probability density is not negligible (e.g., we used $\zeta_{\rm
min} = -15$ and $\zeta_{\rm max} = 10$ for computing $H(t)$ for
$\kappa = 1$).

\end{document}